\documentclass[lineno]{iopart}
\usepackage{multirow}
\usepackage[normalem]{ulem}	
\usepackage{xspace}
\usepackage{xcolor}
\usepackage{graphicx, caption}
\usepackage{lineno,hyperref}

\expandafter\let\csname equation*\endcsname\relax

\expandafter\let\csname endequation*\endcsname\relax
\usepackage{amsmath}
\usepackage{amssymb}
\usepackage{aas_macros}

\usepackage[section]{placeins}

\newcommand{\hasasia}{\texttt{hasasia}\xspace}
\newcommand{\fstat}{$\mathcal{F}_e$ statistic\xspace}

\def\be{\begin{equation}}
\def\ee{\end{equation}}
\def\bi{\begin{itemize}}
\def\ei{\end{itemize}}
\def\ben{\begin{enumerate}}
\def\een{\end{enumerate}}

\begin{document}

\title[Baier et al. 2024]{A sensitivity curve approach to tuning a pulsar timing array in the detection era}




\author[0000-0002-4972-1525]{Jeremy G. Baier}
\address{Oregon State University, 1500 SW Jefferson Ave, Corvallis, OR 97331, United States}
\author[0000-0003-2742-3321]{Jeffrey S. Hazboun}
\address{Oregon State University, 1500 SW Jefferson Ave, Corvallis, OR 97331, United States}
\author[0000-0003-4915-3246]{Joseph D. Romano}
\address{Department of Physics and Astronomy, University of Texas Rio Grande Valley, One West University Boulevard, Brownsville, TX 78520, United States}

\begin{abstract}
    As pulsar timing arrays (PTAs) transition into the detection era of the stochastic gravitational wave background (GWB), it is important for PTA collaborations to review and possibly revise their observing campaigns. The detection of a  “single source” would be a boon for gravitational astrophysics, as such a source would emit gravitational waves for millions of years in the PTA frequency band. Here we present generic methods for studying the effects of various observational strategies, taking advantage of detector sensitivity curves, i.e., noise-averaged, frequency-domain detection statistics. The statistical basis for these methods is presented along with myriad examples of how to tune a detector towards single, deterministic signals or a stochastic background. We demonstrate that trading observations of the worst pulsars for high cadence campaigns on the best pulsars increases sensitivity to single sources at high frequencies while hedging losses in GWB and single source sensitivity at low frequencies. We also find that sky-targeted observing campaigns yield minimal sensitivity improvements compared with other PTA tuning options. Lastly, we show the importance of the uncorrelated half of the GWB, i.e. the pulsar-term, as an increasingly prominent sources of noise and show the impact of this emerging noise source on various PTA configurations.

\end{abstract}

\section{Introduction} 
\label{sec:intro}
Pulsar timing arrays around the world recently published \cite{ipta3p+2024,ng15gwb,eptadr2_3:gwb,pptadr3:gwb} compelling evidence for Hellings--Downs correlations \cite{hd83} for a stochastic gravitational wave background (GWB), ushering pulsar timing arrays (PTAs) into their detection era. This exciting new era brings the expectation of more types of gravitational waves (GWs) to be detected in the near future, including but not limited to resolvable binary supermassive black holes \cite{kelley+2018ill} and GW bursts \cite{finn+2010gwburst} as well as other GW features such as anisotropies in the background \cite{mingarelli+2013_anisotropy,taylor+2013_anisotropy,taylor+2020} and GW memory \cite{madison+2014bwm}. With a tentative background in hand, it is necessary to incorporate this new information -- which may soon be regarded as a foreground with respect to other sources -- and reassess sensitivity to those same sources. This work investigates PTA sensitivity to single sources of GWs as individual binaries emerge above the background. Specifically, we answer the question of how a pulsar timing array can optimize its sensitivity to these single sources in the presence of a GWB.

\subsection{Pulsar timing arrays and pulsar observing campaigns}
PTAs operate by regularly timing a set of galactic millisecond pulsars (MSPs) \cite{foster+1990}. Deviations between the anticipated arrival time and measured arrival time of pulses of light can be correlated across these pulsars to uncover GW signatures in the data \cite{hd83}. A number of collaborations \cite{mclaughlin+2013,manchester+2013ppta,kramer+2013epta} and an international consortium of those collaborations \cite{manchester:2013} have been monitoring more than $100$ MSPs since the early $2000$s in an effort to detect these GWs. Their early efforts mostly established persistent timing campaigns of the lowest residual RMS error pulsars over many years, along with stable telescope equipment that minimized systematic noise.

Once these observing campaigns were well established, the community was able to consider optimizing observations of the galactic population of MSPs in order to detect a particular GW signal. A variety of factors play a role in this type of optimization including which pulsars to observe, how frequently to observe them, and which electromagnetic frequencies to observe them with \cite{lcc+16b,lam+2018opt_freqs}.  Limitations on telescope time and funding constrain PTA collaborations in their optimizations of these pulsar observing campaigns. The two most likely types of sources of GWs sought in the nanohertz band are a stochastic gravitational wave background (GWB) and continuous waves (CWs) from single sources \cite{rosado+2015}. Siemens et al. 2013 \cite{siemens+2013} derives scaling laws for PTAs and finds that for a GWB the dominant factor in signal-to-noise (S/N) growth is the number of pulsars observed as this yields the most pairs of pulsars for Hellings--Downs correlations. On the other hand, various groups \cite{christy+2014, lam18_optcad, liu+2023} found that an optimal campaign for CWs involves allocating more observing time into one's ``best'' pulsars (in this case the ones with the smallest residual RMS errors) -- whether that is longer observations or more frequent observations. In a similar vein, \cite{speri+2023} finds that the majority of a CW signal comes from just a subset of pulsars. 

Most of these optimizations are analytic or semi-analytic in nature, approximating what the most important aspects of the PTA likelihood will be over time. Many other studies, like \cite{speri+2023}, carry out full simulations of pulsar timing arrays with injected GW signals to understand the sensitivity of PTAs to various sources \cite{kelley+2018ill,rosado+2015,taylor+2016there_yet,mingarelli+2017nat,kelley:2018fur} and their ability to estimate GW parameters accurately \cite{liu+2023,pol+2021astro4cast,babak+2024gwb_forecast}. Here we use frequency-domain detector characterization and sensitivity curves built from the noise-averaged expectation value of detection statistics to study not just the time-to-detection of various sources, but also myriad tunable parameters of the observational campaigns that may change the time to detection. These methods mix the best of both worlds, speeding up calculations by using only the expectation value of many noise realizations of a PTA configuration, but also using the frequentist statistics at the heart of many PTA GW searches. All the techniques mentioned above are valuable in their own right; possessing pros and cons for particular applications. Tuning a PTA is the work of sensitivity curves.

\subsection{PTA detector characterization}
Instrument sensitivity is commonly employed to quantify and explore a detector's capacity to differentiate signals of interest from sources of noise. Pulsar timing arrays can be well characterized as detectors by quantifying noise and timing model effects in the pulse arrival times \cite{thrane_romano2013,moore+2015,hazboun:2019sc}.

The software package \hasasia \cite{hazboun:2019has} was developed to robustly and efficiently characterize pulsar timing arrays as detectors. In this paper we introduce additional functionality to \hasasia in order to expand upon techniques that leverage detector characterization as an efficacious tool in both forecasting and optimizing sensitivity to single sources of GWs.

The outline of the paper is as follows. In Section \ref{sec:detstats}, we compare existing detection statistics for continuous GWs emitted by supermassive binary black holes (SMBBHs), elaborating on a polarization, inclination, and phase-marginalized statistic. In Section \ref{sec:pta_sensitivity}, we describe our  model for a PTA similar in characteristics to the International Pulsar Timing Array (IPTA) \cite{manchester:2013}. In Section \ref{sec:detect_prob}, we explore the detection capabilities of a simulated PTA by comparing its sensitivity to simulated SMBBH populations. In Section \ref{sec:tuning_the_detector}, we demonstrate techniques to optimize pulsar observing campaigns for single-source detection and explore how the GWB and a transdimensional simultaneous fit of the GWB and CWs affect single-source sensitivity. Lastly, we conclude in section \ref{sec:discussion} with remarks on how these techniques could be used with a reduced PTA dataset to optimize pulsar timing campaigns for single-source detections and multi-messenger detections. \ref{sec:appendixA} contains a derivation of a marginalized $\mathcal{F}$ statistic that is used throughout the work.

\section{Detection Statistics}
\label{sec:detstats}

In order to forecast the ability of PTAs to detect various GW sources, we use sensitivity curves \cite{thrane_romano2013,moore+2015,hazboun:2019sc,hazboun:2019has} for pulsars based on their noise properties and timing model transmission function to determine the spectra for individual pulsars. These are then combined using two GW frequentist statistics, written in the Fourier domain, to characterize the PTA as a GW detector. Specifically, if one writes down a signal-to-noise ratio, often derived from a maximized likelihood ratio, the denominator, i.e., the ``noise'' as a function of frequency, is taken to be the sensitivity. One such statistic, known as the optimal cross-correlation statistic (OS) \cite{chamberlin+2015os,anholm+2009os}, is used for assessing sensitivity to an isotropic gravitational wave background. The square of the expected signal-to-noise ratio of the optimal statistic is
\be
\rho_{\rm OS}^2=
2T_{\rm obs}\int_0^{f_{\rm Nyq}}{\rm d}f\>
\frac{S_h^2(f)}{S_{\rm eff}^2(f)}
\label{e:rho2_gwb}
\ee
where $S_h$ is the strain power spectral density of the GWB, $T_{\rm obs}$ is the time span of the full PTA dataset and $S_{\rm eff}(f)$ is the {\em effective} strain-noise power spectral density for the whole PTA
\be
S_{\rm eff}(f)
=
\left(\sum_I\sum_{J>I}
\frac{T_{IJ}}{T_{\rm obs}}
\frac{\chi^2_{IJ}}{S_I(f)S_J(f)}
\right)^{-1/2}\,.
\label{e:Seff_gwb}
\ee
Note that $S_{\rm eff}(f)$ can be interpreted as the sensitivity of the PTA to a stochastic GWB. The above expansion includes contributions from the Hellings-Downs factors $\chi_{IJ}$, the overlapping timespan between two pulsar's datasets, $T_{IJ}$, and the individual pulsar strain-noise power spectral densities $S_I(f)$. Note that $S_{\rm eff}(f)$ has dimensions of strain${}^2$/Hz. See \cite{hazboun:2019sc} for more details.  

The other statistic we use is a matched filter statistic for resolvable binary supermassive black holes \cite{hazboun:2019sc,ellis+2012match}. Binary SMBHs are the leading candidate for the source of the GWB seen by PTA collaborations \cite{ng15smbbh}, and is the main focus of this work. As we will demonstrate, tuning a PTA requires not only making it as sensitive as possible to these individual sources, but also maintaining sensitivity to the cross correlations in the GWB. Pulsar observing campaigns can be fine-tuned to maximize sensitivity to the GWB, individual sources, or optimized for both.

There are a few frequentist statistics in the literature developed for single sources in PTAs. See for example \cite{babak+2012, ellis+2012match, ellis+2012fp, taylor+2014Bstat, hazboun:2019sc}. In \ref{sec:appendixA} we review the relationships between these and the matched filter signal-to-noise ratio used for sensitivity curves in this work. The signal-to-noise ratio for a circular, non-evolving binary black hole source can be written as
\be
\begin{aligned}
\rho^2_{\rm MF}(h_0,\iota,\psi,\hat{k})
&= 2 T_{\rm obs}\int_0^{f_{\rm Nyq}} {\rm d}f\>
\frac{S_h(f,h_0)}{S_{\rm eff}(f,\hat k,\iota,\psi)}\,,
\label{e:rho2(k)}
\end{aligned}
\ee
where
\begin{align}
&S_{\rm eff}(f,\hat k, \iota, \psi)\equiv \left(
\frac{4}{5}\sum_I \frac{T_I}{T_{\rm obs}}\frac{1}{S_I(f,\hat k,\iota,\psi)}\right)^{-1}\,.
\label{e:Seff_inc}
\end{align}
The individual pulsar strain-noise power spectral densities are functions of GW frequency $f$, 
source direction $-\hat k$, inclination angle $\iota$, and polarization angle $\psi$.
This sensitivity curve formalism and implementation is robust and has been corroborated with Bayesian methods in \cite{hazboun:2019sc} and \cite{ng15detchar}.

Of particular interest for these frequentist statistics is the ability to quote false alarm probabilities given a null distribution and detection probabilities given a distribution of the noise with a signal in it. The (null) distribution of the optimal statistic follows a generalized $\chi^2$ distribution \cite{hazboun+2023gx2} and can be calculated for a given dataset given a set of noise parameters. This work focuses on single-source detections with the assumption that we are far enough in the future to assume a significant detection of the GWB. 

In \ref{sec:appendixA} we introduce the \fstat precisely because it has a well-defined false alarm probability/detection probability and can be related to $\rho^2_{\rm MF}$ in Equation~\ref{e:rho2(k)}. 
For our purposes, the main result of  \ref{sec:appendixA} is demonstrated in Figure~\ref{fig:DP_compare} where we motivate the use of an $\iota$, $\psi$ and $\phi_0$-averaged detection probability, $\left\langle {\rm DP}_{\mathcal{F}_e}(\rho(\iota,\psi)) \right\rangle_{\iota, \psi, \phi_0}$, based on the fact that 2 times the \fstat follows a noncentral $\chi^2$ distribution \cite{babak+2012, ellis+2012fp}. Therefore, we can calculate the detection probability given a particular value of $\rho^2_{\rm MF}(h_0,\iota, \psi, \hat k)$. As we will see in Section ~\ref{sec:detect_prob}, this allows us to use our sensitivity curves to efficiently carry out the analyses like those in \cite{kelley+2018ill,rosado+2015,gardiner+2024}.

\section{PTA Sensitivity}
\label{sec:pta_sensitivity}

\subsection{Simulating an IPTA-like Detector}
To demonstrate our methods, we simulate a detector which is similar in its number of pulsars and time span to the current IPTA \cite{ipta3p+2024} dataset and similar in its noise characteristics to the NANOGrav $15$ year data set \cite{ng15data}. 

\begin{figure}[ht]
    \centering
    \includegraphics[width=0.95\columnwidth]{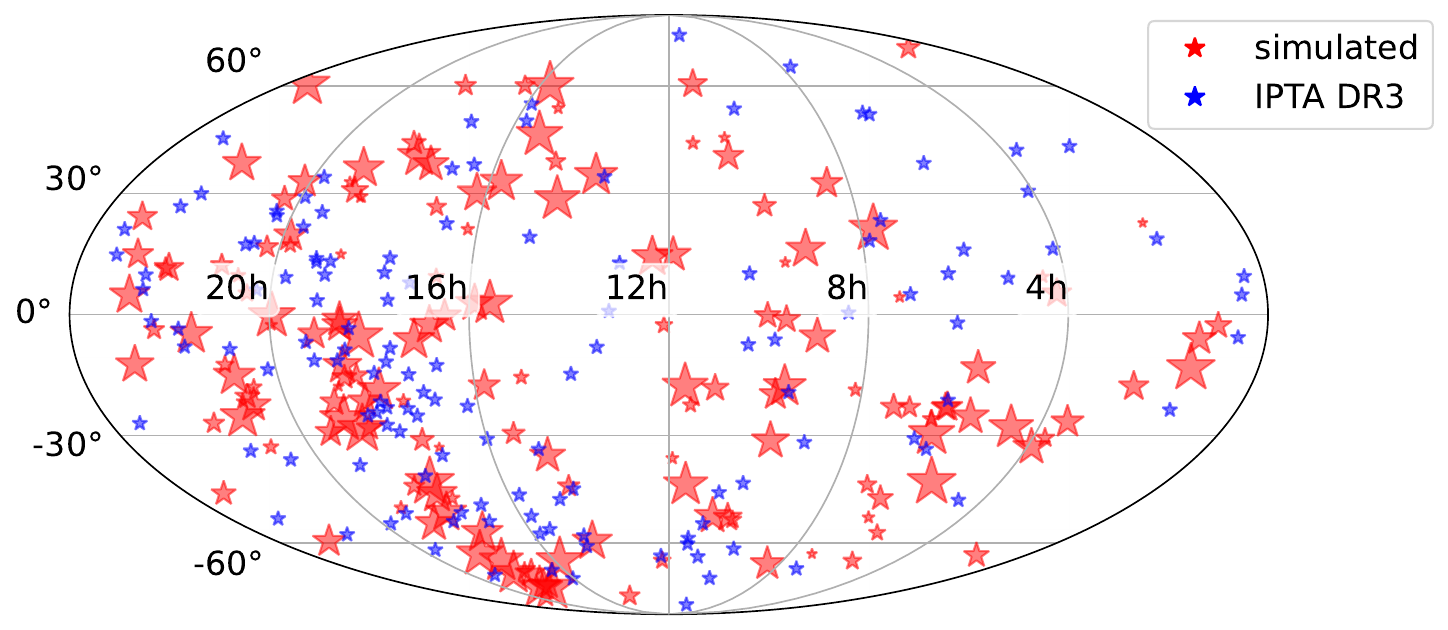}
    \captionsetup{width=0.85\linewidth}
    \caption{Pulsar sky locations. IPTA DR3 pulsars are marked by a blue star and pulsars in our simulated dataset are marked by a red star. The red stars are sized according to their individual pulsar strain-noise power spectral densities, $S_{I}(f)$ (see Equation~\ref{e:Seff_gwb}), where a larger star corresponds to a larger value of $S_I(f)$. \label{fig:sky_locations}}
\end{figure}

Using the approximate sky locations of the pulsars in \cite{ipta3p+2024}, we create empirical distributions of pulsar ecliptic longitude and latitude from which we initially draw $115$ pulsars, which are shown in Figure \ref{fig:sky_locations}.

Similarly, we construct an empirical distribution of the logarithm of pulsar white noise reported in \cite{ng15data}, using the ``whitened'' values in cases where red noise is found to be significant.
Next, we create a 2--dimensional empirical distribution of the amplitudes and spectral indices of the power-law model of achromatic red noise used in \cite{ng15detchar},
\be
\varphi_i \sim \frac{A^2}{12\pi^2T_{\text{span}}}\left(\frac{f_i}{f_\text{yr}}\right)^{-\gamma}\text{yr}^3.
\ee
We construct this distribution from the subset of pulsars whose intrinsic red noise remained significant in a common red noise search. (See the bolded values in Table $2$ of \cite{ng15detchar}.) This allows us to later inject a common red noise process without double counting red noise in the pulsars. Since only $\sim18\%$ of the pulsars in the NANOGrav $15$ year dataset maintain significant levels of red noise after removing the common red noise signal, we only inject intrinsic red noise in a proportional number of pulsars in our simulated dataset. Lastly, we construct an empirical distribution of pulsar time spans from \cite{ng15data} to model historical PTA growth.

From the constructed empirical distributions for pulsar sky coordinates, white noise, red noise, and time span, we simulate a pulsar timing array of $115$ pulsars with a total time span of $16$ years and an observing cadence of $20$ observations per year for all pulsars. We simulate these data with the \hasasia package and proceed to inject into all pulsars a power-law gravitational wave background, which we construct from the median of the joint posterior from \cite{ipta3p+2024}: $\log_{10}A_{\rm GWB} =-14.29$ and $\gamma_{\rm GWB} =3.44$. From here, we adjust the overall sensitivity of our detector to mirror that of contemporary PTAs by adjusting the overall power in the white noise of the detector, given by, 
\be
P_{\text{WN}} = 2\sigma^2/c\,,
\ee
where $c$ is the cadence of observations and $\sigma$ is the uncertainty on pulse arrival time. By shifting our white noise distribution for all the pulsars, we adjust the overall sensitivity until the injected GWB has a S/N of $~7$ at $16$ years to match preliminary projects for IPTA Data Release 3 (DR3) \cite{ipta3p+2024}. In other words, we are calibrating our PTA to have a sensitivity at $16$ years, which is comparable to the projections of DR3.

\begin{figure}[hbt!]
    \centering
    \includegraphics[width=0.85\textwidth]{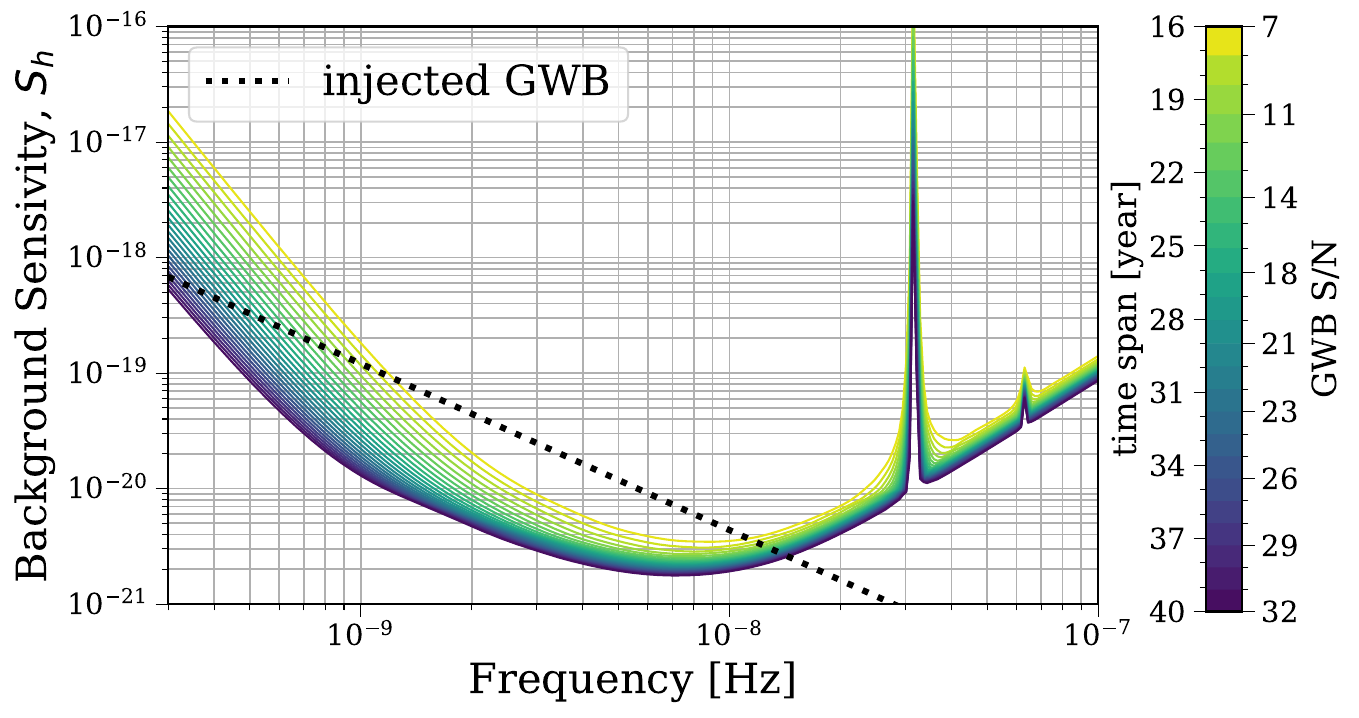}
    \captionsetup{width=0.85\linewidth}
    \caption{Standard GWB sensitivity. The background sensitivity curves of the simulated PTA are plotted for every year between $16$ and $40$ years. The color bar encodes both the year and the GWB S/N at that year. The injected GWB is shown as a dashed black line.}
    \label{fig:pta_tuning}
\end{figure}

\begin{figure}[hbt!]
    \centering
    \includegraphics[width=0.85\textwidth]{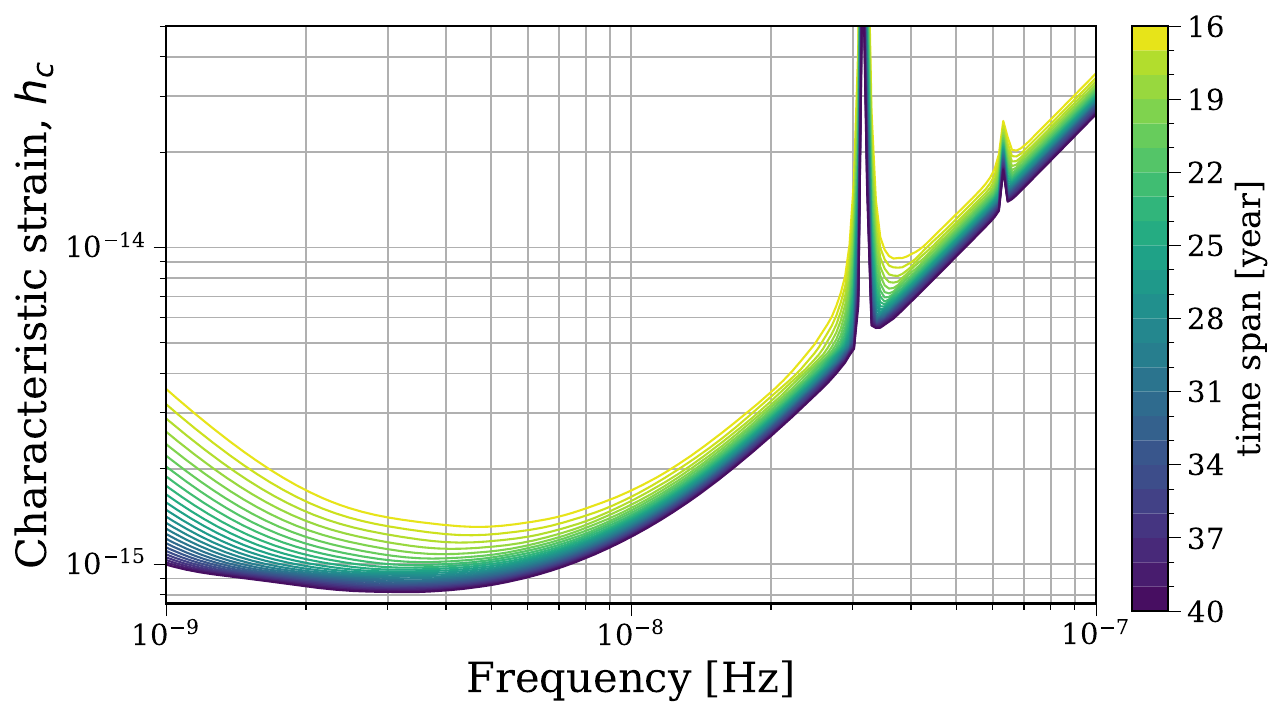}
    \captionsetup{width=0.85\linewidth}
    \caption{Standard single source sensitivity. The single-source sensitivity curves of the simulated PTA are plotted for every year between $16$ and $40$ years. The color bar encodes the dataset timespan for which the sensitivity curve is calculated.}
    \label{fig:pta_tuning_hc}
\end{figure}

From here, we simulate our detector out to a total time span of $40$ years. We assume that the white noise in our detector does not change over the course of the simulations (i.e., the radio telescopes or other noise mitigation techniques do not improve), and we add $2$ pulsars per year whose sky locations and noise properties are drawn from the same empirical distributions used above. As is standard, we only include pulsars in our dataset or a time slice of the data set if the pulsar has at least 3 years of data. Generally, PTAs do this so that a timing model may be accurately fit. So at $40$ years, we have a detector with $156$ pulsars and a GWB S/N of $\sim32$. The time evolution of the GWB sensitivity is shown in Figure \ref{fig:pta_tuning}, and the time evolution of the sky-averaged single-source characteristic strain sensitivity is shown in Figure \ref{fig:pta_tuning_hc}. Note that the main purpose of this PTA is to be a high fidelity test bed in order to demonstrate the various techniques developed for helping to inform observing era strategies for pulsar timing. A large scale effort is underway to forecast the sensitivity of PTAs using next generation facilities. 

\section{Single SMBBH Detection Prospects from a GWB}
\label{sec:detect_prob}

One way we can assess the single source detection capacity of our simulated detector is by computing the detection statistics from \cite{rosado+2015} (see Section 4.2 for a brief review) over a simulated population of SMBBHs. By looking at many realizations of these populations, we can make statistical statements about how many binaries we detect, when we detect them, and at which frequencies we detect them.

\subsection{SMBBH Population Simulations}
In this study, we use $1000$ realizations of a SMBBH population simulated with \texttt{holodeck} \cite{holodeck2023}, a software package designed for astrophysical inference of black hole population models using PTA data. See \cite{ng15smbbh}, \cite{gardiner+2024} for a description. 
As demonstrated in \cite{kelley+2018ill,gardiner+2024} populations of this type vary widely between choices of the black hole-population parameter space and also from realization to realization. Therefore, we use $10$ random samples from the same phenomenological parameter space in \cite{ng15smbbh}, but for each sample of the astrophysical black hole-population parameters, we generate $100$ population realizations which we then adjust by ``tilting'' the median background to match the 3P+ spectral index, $\gamma_{\rm GWB} =3.44$. In order to minimize distractions in this methods paper from the huge diversity in these binary populations, which are \textbf{not} conditioned on any PTA data, we choose a set of realizations that gave us a ``middle of the road'' total detection probability (see Section~\ref{s:det_stats} below) median of roughly $\sim50$\% at 30 years. This population is merely a control from which to measure \emph{changes} in the detection statistics throughout the various optimizations we discuss in this work.

\begin{figure}[ht]
    \centering
    \includegraphics[width=0.85\textwidth]{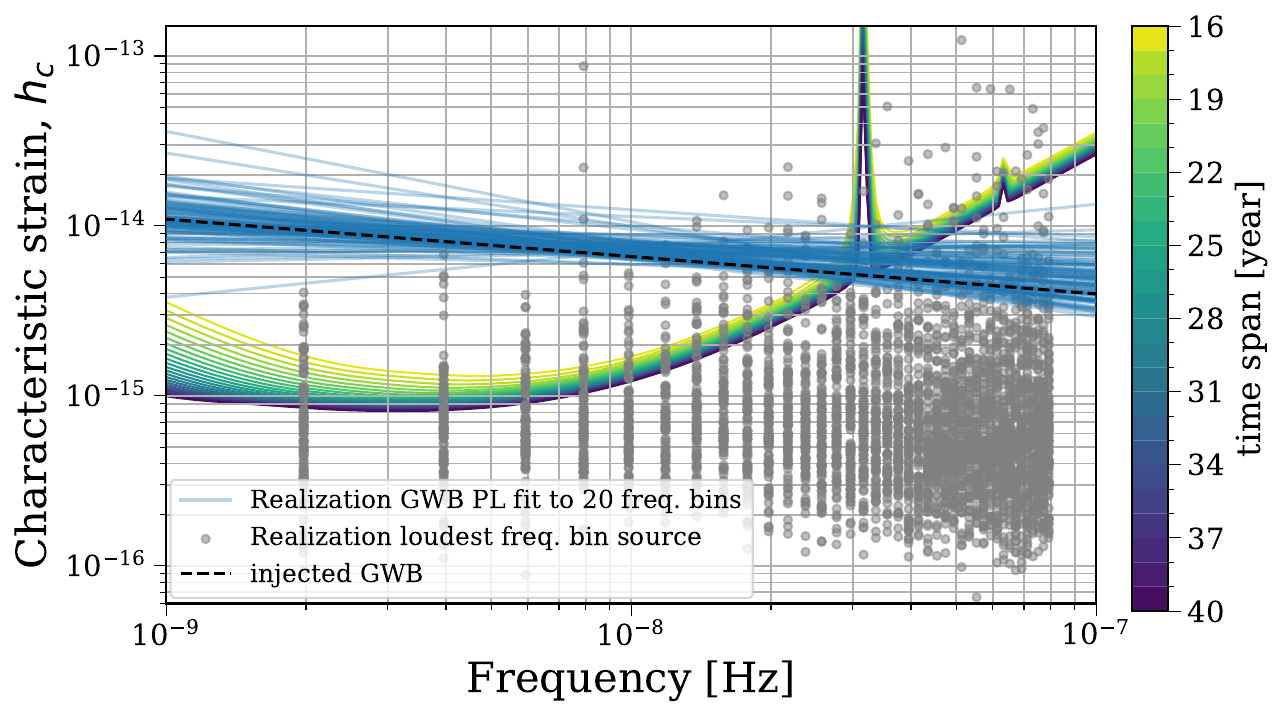}
    \captionsetup{width=0.85\linewidth}
    \caption{Simulated binary population sourcing the GWB. The loudest strain in each frequency bin for $100$ of the $1000$ total SMBBH population realizations compared to a time-evolving detector sensitivity. As the data set time span increases, the binaries grow in S/N, some eventually reaching a detection threshold.}
    \label{fig:rosado1}
\end{figure}

 Recall that in the course of simulating our detector, we injected a power-law GWB into all the pulsars which manifests in the detector sensitivity as a GWB self-noise. Here we note that we leave our detector sensitivity fixed in this regard, electing to not recompute detector sensitivity for specific GWB realizations since we have ensured that all the population realizations are broadly consistent with the injected background. We leave an investigation of the effects of power-law deviations on detector sensitivity as future work. Each realization contains binaries in $40$ frequency bins evenly spaced from $\sim2$nHz to $\sim80$nHz. We consider detection statistics on the loudest (i.e. the largest strain amplitude) SMBBH in each frequency bin. Figure~\ref{fig:rosado1} shows the characteristic strain ($h_c$) of the loudest binary in each frequency bin for all $1000$ realizations considered.

We focus on how one can harness these sensitivity curves as an efficient alternative to costly simulations, allowing us to easily adjust the signal(s) and noise in our detector to optimize the myriad knobs for single-source detections. In future work, we will use \texttt{holodeck} to generate many SMBBH populations which have been conditioned on the free spectral posteriors from \cite{ng15gwb}. This more in-depth treatment should yield more robust time-to-detection predictions. In this work, we present the methods developed for such studies.

\subsection{Application of detection statistics to population realizations}\label{s:det_stats}

\begin{figure}[ht]
    \centering
    \includegraphics[width=0.85\textwidth]{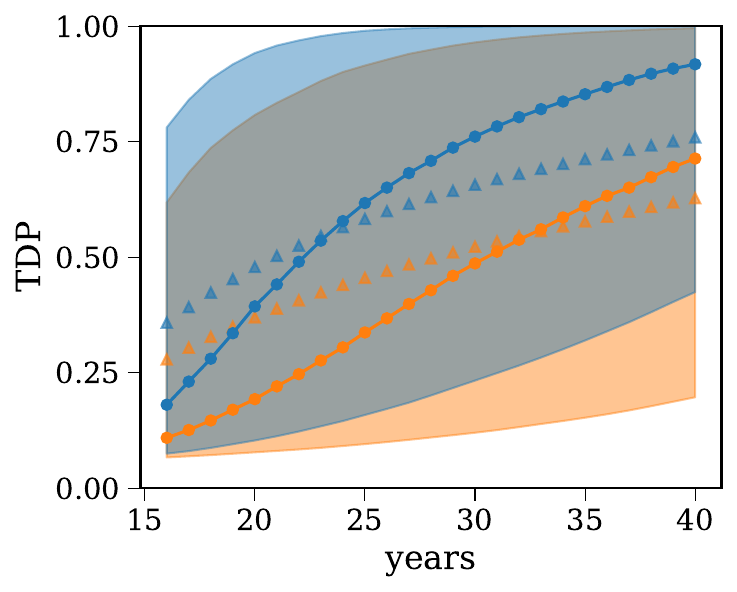}
    \captionsetup{width=0.85\linewidth}
    \caption{Total detection probability of the standard observing campaign. Both the mean (triangles) and the median (dots) TDP across realizations are shown at each year from years $16$ to $40$ for the simulated detector. They are calculated both with (blue) and without (orange) a GWB Earth-term simultaneous fit. (See Section \ref{sec:tuning_the_detector}.4 for details on a simultaneous fit.) The shaded regions enclose the middle $68\%$ of the TDP distribution.}
    \label{fig:total_dp_evolution}
\end{figure}

There are a number of ways one could apply the detection statistics from \cite{rosado+2015}  for these SMBBH population realizations using the matched-filter (squared) S/N $\rho_{\rm MF}^2$ described in Section~\ref{sec:detstats}. We consider two useful approaches: a per--frequency bin detection probability as well as a total detection probability. 
The per--frequency bin S/N calculates the S/N described in Section \ref{sec:detstats} and \ref{sec:appendixA} for the loudest source in each frequency bin across all $1000$ population realizations.
A per--frequency detection probability, $\text{DP}_i$, is calculated using the probability density function of the \fstat given a value for the S/N ($\rho$), see for example Equation $32$ from \cite{rosado+2015} and the discussion in \ref{sec:appendixA} surrounding Equation~\ref{eq:dp_P}.
The total detection probability (TDP) is then given by (Equation $33$ in \cite{rosado+2015}),
\be
\text{TDP}=1-\prod_i[1-\text{DP}_i],
\ee
where $i$ indexes frequency bins. The TDP can be interpreted as the probability of detecting at least one binary in \textit{any} frequency bin.
For a ``$3\sigma$'' detection, we use the false alarm rate to be $0.00135$, which yields $\bar{\mathcal{F}_e}\sim8.89$, requiring a S/N of $\sim3.7$ under this detection statistic's null distribution which is a non-central $\chi^2$ distribution with $4$ degrees of freedom and a non-centrality parameter of $\rho^2$. For the entirety of this work, we use this detection threshold of $\bar{\mathcal{F}_e}\sim8.89$. We apply these detection statistics to the simulated population realizations and the simulated PTA described in Section~\ref{sec:pta_sensitivity}.
Firstly, we compute the TDP for each realization of the universe every year from $16$ to $40$ years in Figure \ref{fig:total_dp_evolution}. Again, this probability can be thought of as the probability of making a detection of \textit{at least} one single source in \textit{any} of the frequency bins. We note that this is asymptotic to ${\rm TDP|_{\rm DP_i=FAP}}~\sim0.05$ at low probabilities since for a given non-zero false alarm probability, there is a non-zero chance of falsely claiming a detection in each frequency bin. Figure~\ref{fig:total_dp_evolution} shows the change in the TDP over the dataset timespan, by tracking the evolution of the mean, median, and the central $68\%$ region.

\begin{figure}[ht]
    \centering
    \includegraphics[width=0.85\textwidth]{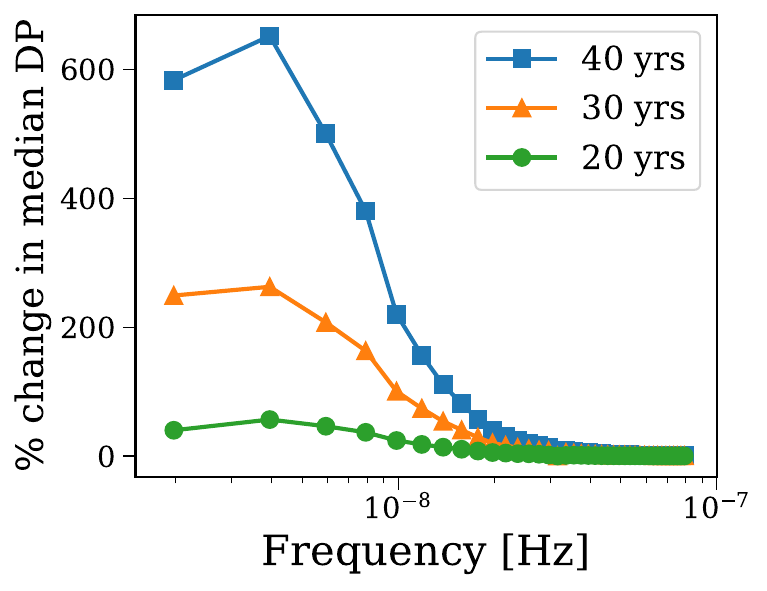}
    \captionsetup{width=0.85\linewidth}
    \caption{The percent change in median detection probability. The $\%\Delta$DP is shown from year $16$ to years $20$, $30$, and $40$ is plotted for the $40$ frequency bins considered in this study.} 
    \label{fig:per_freq_median_dp_evolution}
\end{figure}

Figure~\ref{fig:per_freq_median_dp_evolution} shows how the per--frequency DP$_i$ changes in time by plotting the percent change in the median DP$_i$ from a dataset timespan of $16$ years to $20$, $30$, and $40$ years. Here and throughout the work we define the percent change in a quantity as
\be 
100\times\left(\frac{\mathrm{New}-\mathrm{Original}}{\mathrm{Original}}\right).
\ee
Figure~\ref{fig:per_freq_median_dp_evolution} reveals how low frequency binaries tend to move to higher detection probabilities more quickly than high frequency binaries for our simulated populations and PTA, which is expected since the sensitivity at low frequencies is primarily a function of the timespan of the data set.


We return to this methodology of comparing sensitivity to simulated populations in the next section, where we optimize our pulsar observing campaigns to single-source detections.

\section{Tuning the Detector}
\label{sec:tuning_the_detector}
A number of substantial changes can be made to pulsar observing campaigns, i.e. observing some pulsars more or less frequently, or with different instruments, or not observing them at all, with the goal of optimizing a PTA's dataset (and thus sensitivity) to a specific source. By using sensitivity curves and frequency-domain detection statistics, we can explore how these changes in a campaign will affect sensitivity without having to re-simulate entire datasets. This efficient approach, paired with the detection statistics described in Sections~\ref{sec:detstats} and \ref{sec:detect_prob}, enables us to robustly quantify our sensitivity to the GWB and the detectability of single sources under a variety of campaigns and noise assumptions. In this section, we highlight our ability to answer the question of how changes to pulsar observing campaigns impact single-source sensitivity.  For the rest of this section, we compare the standard, status quo, campaign (see Figure~\ref{fig:pta_tuning_hc}) to ``alternate'' campaigns at $30$ years assuming that these changes were initiated at $20$ years, putting the alternate campaigns in effect for a total of $10$ years, which is $1/3$ of the considered data set time span. We settle on an optimal campaign for single-source detections under realistic observing constraints and proceed to use the techniques from Section \ref{sec:detect_prob} to assess how detection prospects improve under this optimal campaign. We proceed to explore other topics beyond simply single source sensitivity, such as a simultaneous fit of the GWB alongside single sources, the dependence of sensitivity on the GWB spectrum itself, multi-messenger detection prospects, and sensitivity to new-physics models.

\subsection{High Cadence Campaigns}\label{sec:baseline}
\begin{figure}[ht]
    \centering
    \includegraphics[width=0.85\textwidth]{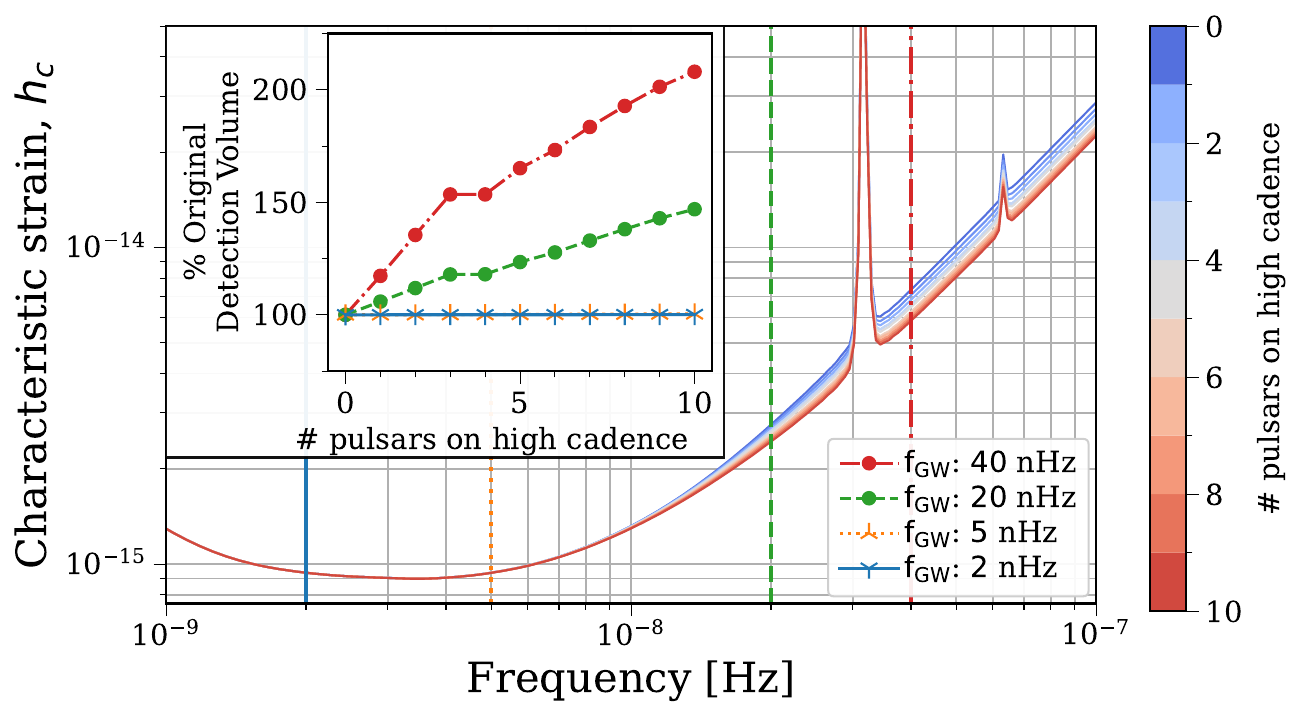}
    \captionsetup{width=0.85\linewidth}
    \caption{Adding observations. Sky-marginalized single-source sensitivity curves are plotted for a variable number of high cadence campaigns, which are encoded in the color bar. The vertical lines denote frequencies at which we compute the change in detector volume with respect to the standard campaign. The change in detector volume  as a function of the number of pulsars on high cadence campaigns is shown in the inset plot.
    }
    \label{fig:alt_camp_x4_cad_sc}
\end{figure}

To start, we simulate a hypothetical observing campaign in which we add high cadence campaigns on our ``most sensitive'' $N$ pulsars.
These high cadence campaigns consist of timing these pulsars four times as frequently from year $20$ to year $30$. For the rest of the work, we define the ``most sensitive'' pulsars as those which have the smallest residual root-mean-square (RMS) error. More frequent observations of these pulsars increase the sensitivity at high frequencies, where white noise dominates in PTAs. High cadence campaigns on $10$ pulsars constitutes a $30\%$ increase in telescope time beyond the assumed yearly increases in telescope time to add $2$ pulsars per year.
As expected, Figure~\ref{fig:alt_camp_x4_cad_sc} shows large sensitivity gains at high frequencies but leaves the sensitivity relatively unaffected at low frequencies, where sensitivity is primarily a function of the timespan of the dataset. The GWB S/N increases $<5\%$ with all $12$ pulsars on high cadence campaigns. Another astrophysically useful way to quantify the sensitivity in our detector is by computing its detection volume (DV). We define the detection volume at a given frequency as
\be
\rm DV(f)
=
\frac{1}{3}\iint\,dr \,d\Omega\,r^2 ,
\label{e:det_vol}
\ee
where $\Omega$ is the solid angle on the sky in steradians and $r$ is the maximum luminosity distance at which the detector is capable of making a $3\sigma$ detection to a binary with a chirp mass of $10^9 \rm M_{\odot}$. Note that our choice in chirp mass and significance threshold are both inconsequential as we quote the percent original DV, which we define as
\be
\%\,\mathrm{Original\,DV}
=
100\times (\mathrm{New \,DV} / \mathrm{Original \,DV}).
\label{e:det_vol_change}
\ee
The inset plot in Figure~\ref{fig:alt_camp_x4_cad_sc} shows that the detector volume remains essentially the same at low frequency and nearly doubles at high frequency, which follows our gains in sensitivity. The hypothetical observing campaign examined in this section provides a useful baseline for our more realistic fixed observing-time campaigns.
%

\subsection{Comparison of Fixed Observing-Time Campaigns}\label{sec:opt}
It is more realistic to consider a scenario where a PTA has a fixed amount of observing time and can only optimize its sensitivity to single sources by reallocating that time amongst pulsars and observations. For this study, we assume a fixed integration time for each observation (i.e. each observation is the same duration) but vary the cadence of observations, which we define as the number of observations of a pulsar per year.

\begin{figure}[ht]
    \centering
    \includegraphics[width=0.85\textwidth]{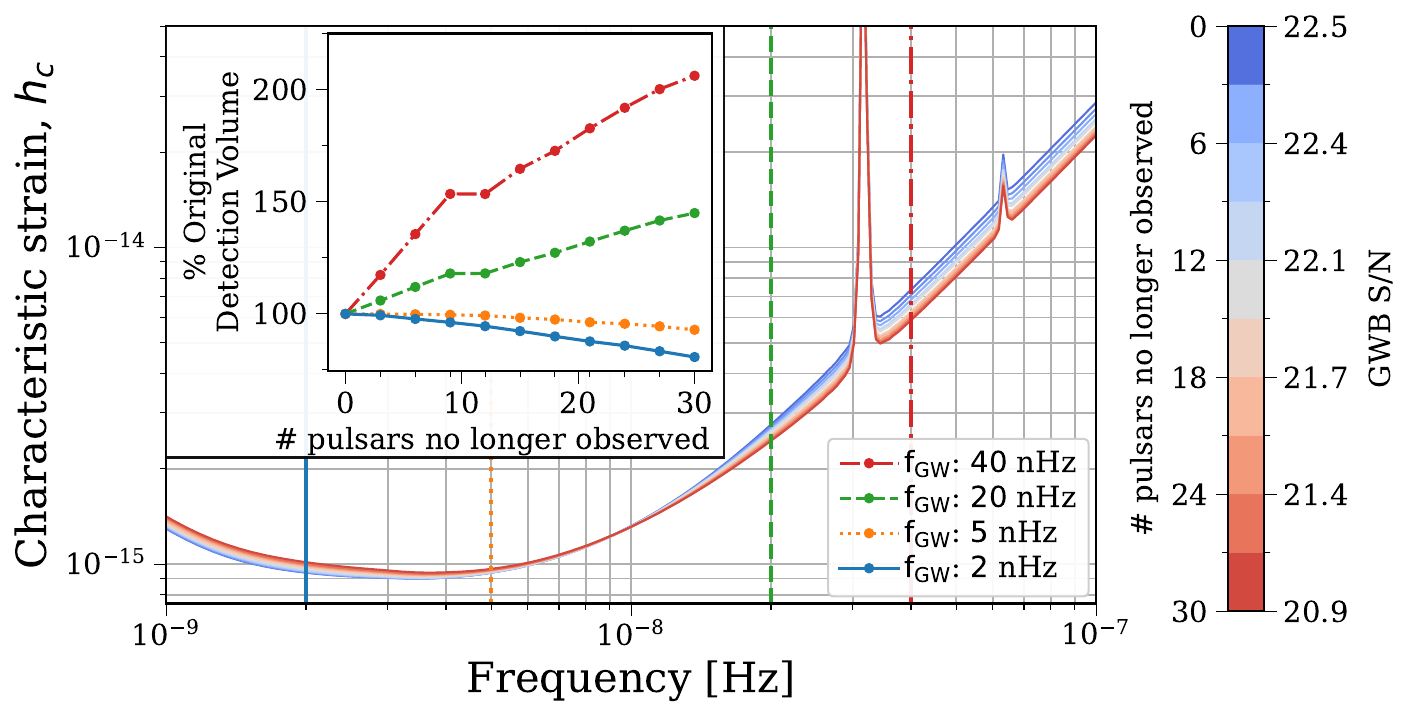}
    \captionsetup{width=0.85\linewidth}
    \caption{Dropping pulsars.
    Sky-marginalized single-source sensitivity curves are plotted for a variable number of pulsars dropped at $20$ years to free telescope time for high cadence campaigns on the most sensitive pulsars. The color bar encodes the number of pulsars dropped (in sets of $3$ pulsars) as well as the GWB S/N. The inset plot tracks the change in detector volume at $4$ different reference frequencies as a function of changes to the observing campaign.}
    \label{fig:x4cad_gwb_psr_drop}
\end{figure}

First, we consider a campaign which stops observing the least sensitive $3N$ pulsars at $20$ years and instead observes the most sensitive $N$ pulsars with a high cadence campaign of quadrupled observations, ensuring that telescope time remains fixed. We test various metrics for ``least sensitive'' pulsar, including largest residual RMS error, but we ultimately use the pulsar's contribution to the GWB. Accordingly, we drop pulsars (in sets of $3$) which contribute the least to the GWB as calculated by leaving out pulsars one-by-one and recomputing the GWB S/N. Again, we define the most sensitive pulsars as those with the smallest residual RMS error. The single-source sensitivity curves in Figure \ref{fig:x4cad_gwb_psr_drop} illustrate how discontinuing observations on the least sensitive pulsars and quadrupling the cadence of the most sensitive pulsars in a fixed-observing timing scenario yields sensitivity gains at high frequencies and sensitivity losses at low frequencies. This sensitivity loss is an important conclusion of this work. The sensitivity of PTAs to single sources at low frequencies is critically dependent on the ability of the detector to mitigate the effects of the GWB, which we explore further in Section~\ref{sec:gwb_dependance}. The change in detector volume is nearly identical to the previous campaign (Figure~\ref{fig:alt_camp_x4_cad_sc}) at high frequencies, but we note a decrease of up to $10\%$ in detector volume at low frequencies due to the discontinuation of observations on pulsars which were contributing to the GWB mitigation.

\begin{figure}[ht]
    \centering
    \includegraphics[width=0.85\textwidth]{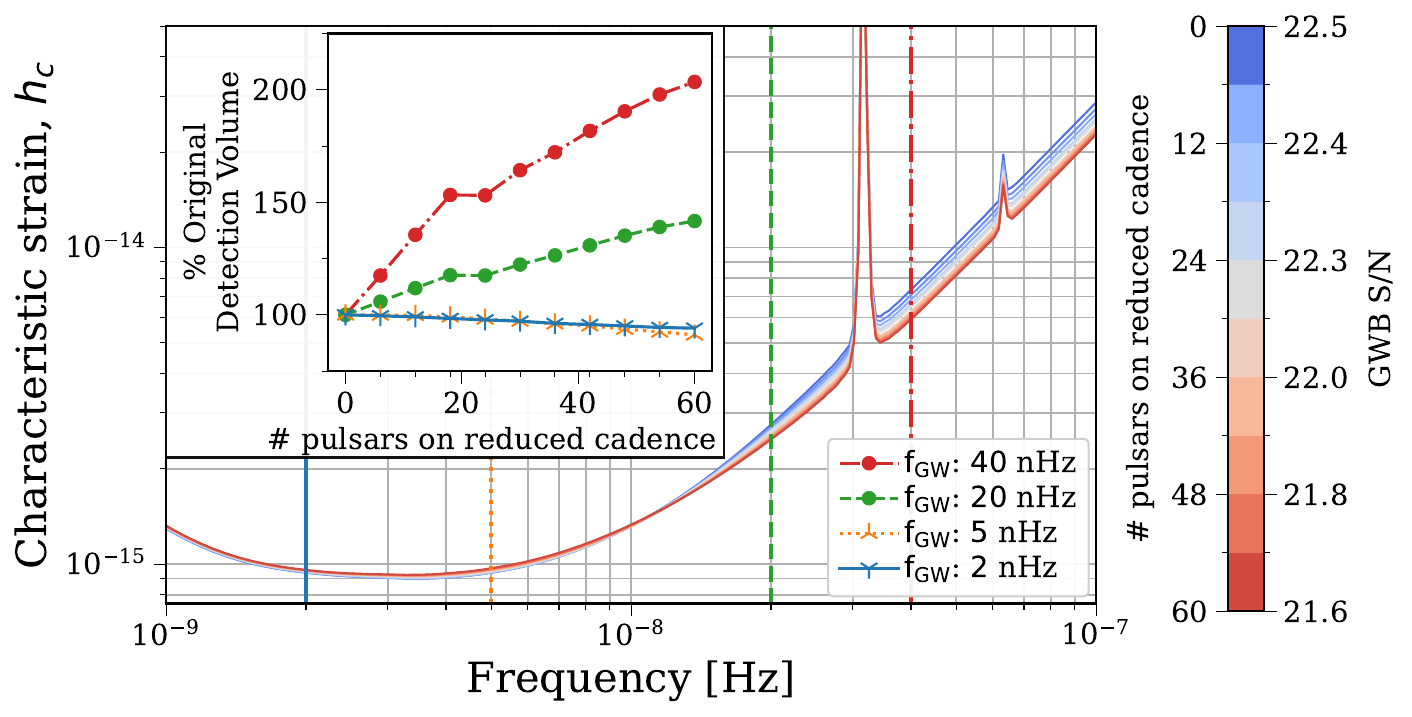}
    \captionsetup{width=0.85\linewidth}
    \caption{Shifting observations. Sky-marginalized single-source sensitivity curves are plotted for a variable number of pulsars whose cadence gets halved at $20$ years to free telescope time for high cadence campaigns on the most sensitive pulsars. The color bar encodes the number of pulsars on low cadence (in sets of $6$ pulsars) as well as the GWB S/N. The inset plot tracks the change in detector volume at $4$ different frequencies as a function of changes to the observing campaign.}
    \label{fig:alt_camp_half_cad_sc}
\end{figure}

\begin{figure}[ht]
    \centering
    \includegraphics[width=0.85\textwidth]{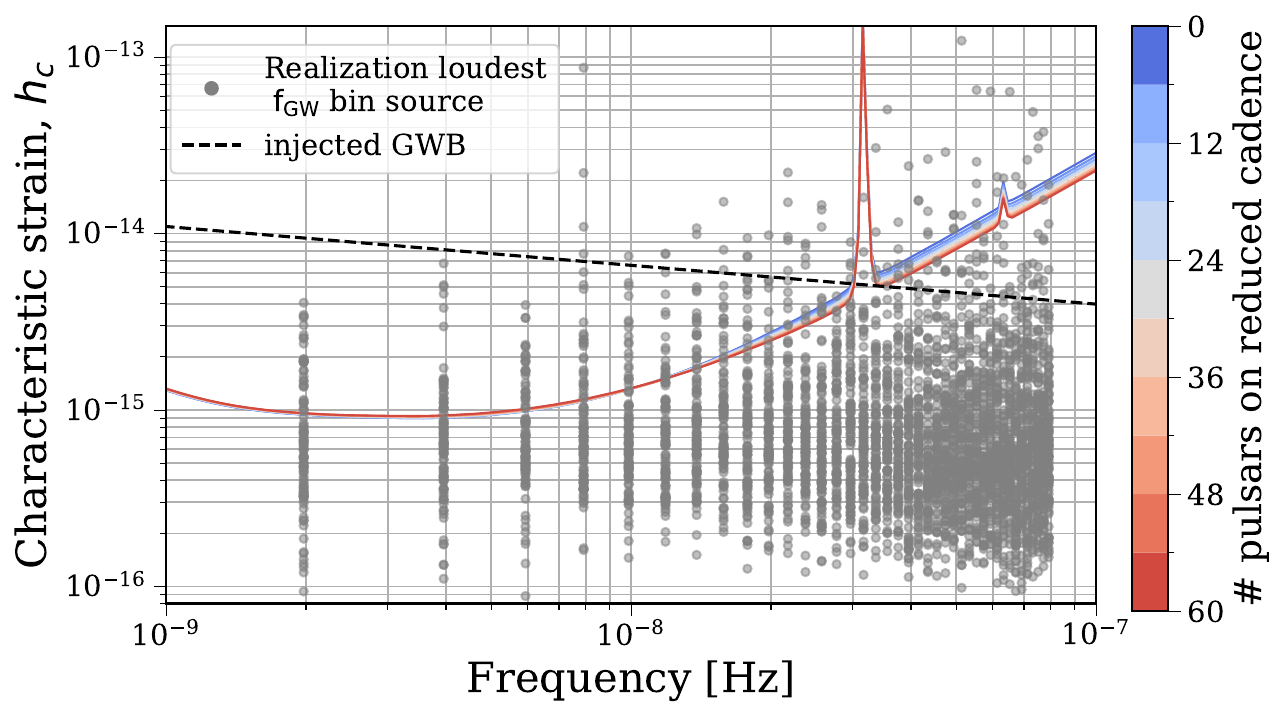}
    \captionsetup{width=0.85\linewidth}
    \caption{The loudest strain in each frequency bin across $100$ population realizations plotted next to the sky-marginalized, single-source strain sensitivity of the detector reveals which binaries might be detectable.}
    \label{fig:alt_camp_rosado_sc}
\end{figure}

\begin{figure}[ht]
    \centering
    \includegraphics[width=0.4\textwidth]{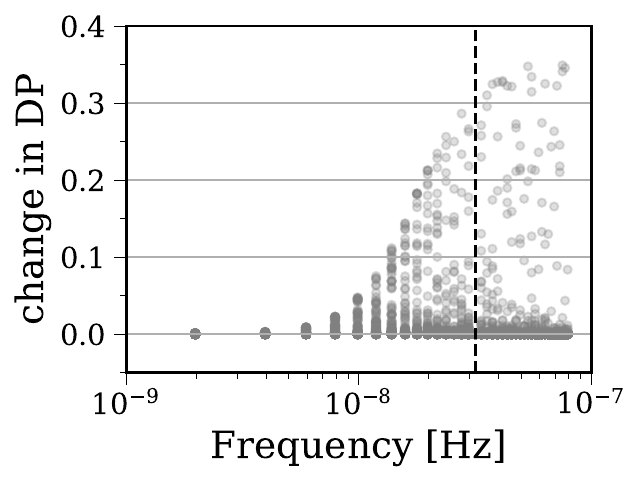}
    \includegraphics[width=0.4\textwidth]{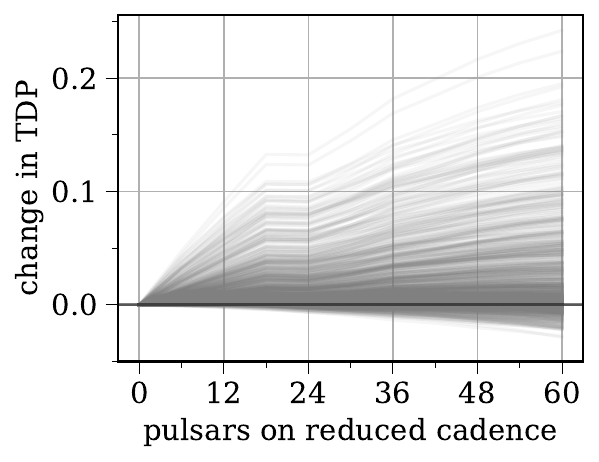}
    \captionsetup{width=.8\linewidth}
    \caption{Changes in detection statistics of the optimal vs. the standard campaign. (a) The change in detection probability between the standard campaign and the optimal campaign with $60$ ($10$) pulsars on reduced (high) cadence is plotted for each frequency bin across realizations. The green line partitions the PTA frequency band into a low and high GW frequency sub bands, the latter of which we might expect to see more multi-messenger candidates. This is further explored in Section~\ref{sec:mma}. (b) The change in total detection probability is graphed as a function of change in observing campaign for the optimal campaign.}
    \label{fig:opt_camp_double_fig_rosado}
\end{figure}

We now consider a campaign which puts the least sensitive $6N$--pulsars on half-cadence while keeping the most sensitive $N$--pulsars on high cadence campaigns, which again keeps the total telescope time fixed. This strategy allows for the continued contribution of the least sensitive pulsars at low frequencies but still frees up telescope time for high cadence campaigns on the most sensitive pulsars. Figure \ref{fig:alt_camp_half_cad_sc} reveals how this campaign, compared to the previous campaign (Figure~\ref{fig:x4cad_gwb_psr_drop}), results in a higher GWB S/N at $30$ years, less sensitivity losses at low frequencies, and roughly the same sensitivity gains at higher frequencies. Accordingly, the detector volume losses for low frequencies are reduced from $10\%$ to $< 5\%$ while the increases in detector volume are maintained. In this way, we are able to preserve the gains of the previous campaign and minimize the losses at low frequencies, all while still keeping telescope time fixed; for this reason, we refer to this campaign as the ``optimized'' campaign for the rest of the work. Figure~\ref{fig:alt_camp_rosado_sc} visualizes what effect this type of optimal campaign has on detecting individual binaries by plotting the loudest strain in each frequency bin for a sample of $100$ of the total $1000$ realizations.
We proceed with this optimized campaign for single-source sensitivity and employ the functionality demonstrated in Section \ref{sec:detect_prob} to further probe these sensitivity gains. Similar to before, we calculate the per-frequency DP$_i$ and the TDP, but now we plot the change in DP$_i$ and the change in TDP between the standard campaign and the optimal campaign. Figure~\ref{fig:opt_camp_double_fig_rosado}(a) illustrates how sensitivity gains at high frequencies correspond to increases in the detection probabilities of individual binaries at those frequencies while we see a decrease in detection probability at lower frequencies. We see the overall effect that the optimal campaign has on detectability in Figure~\ref{fig:opt_camp_double_fig_rosado}(b), which traces TDP as a function of the number of pulsars on low cadence. Most of the population realizations exhibit sizable increases in TDP with more pulsars being moved to the optimal campaign, however some fraction of the realizations -- those corresponding to a population-loudest binary at low frequencies -- exhibit a small decrease in TDP at $30$ years with this campaign due the slight losses in sensitivity at those frequencies.

\subsection{Sky-Targeted High Cadence Campaigns}\label{sec:sky_target}
\begin{figure*}[ht]
    \centering
    \includegraphics[width=0.98\textwidth]{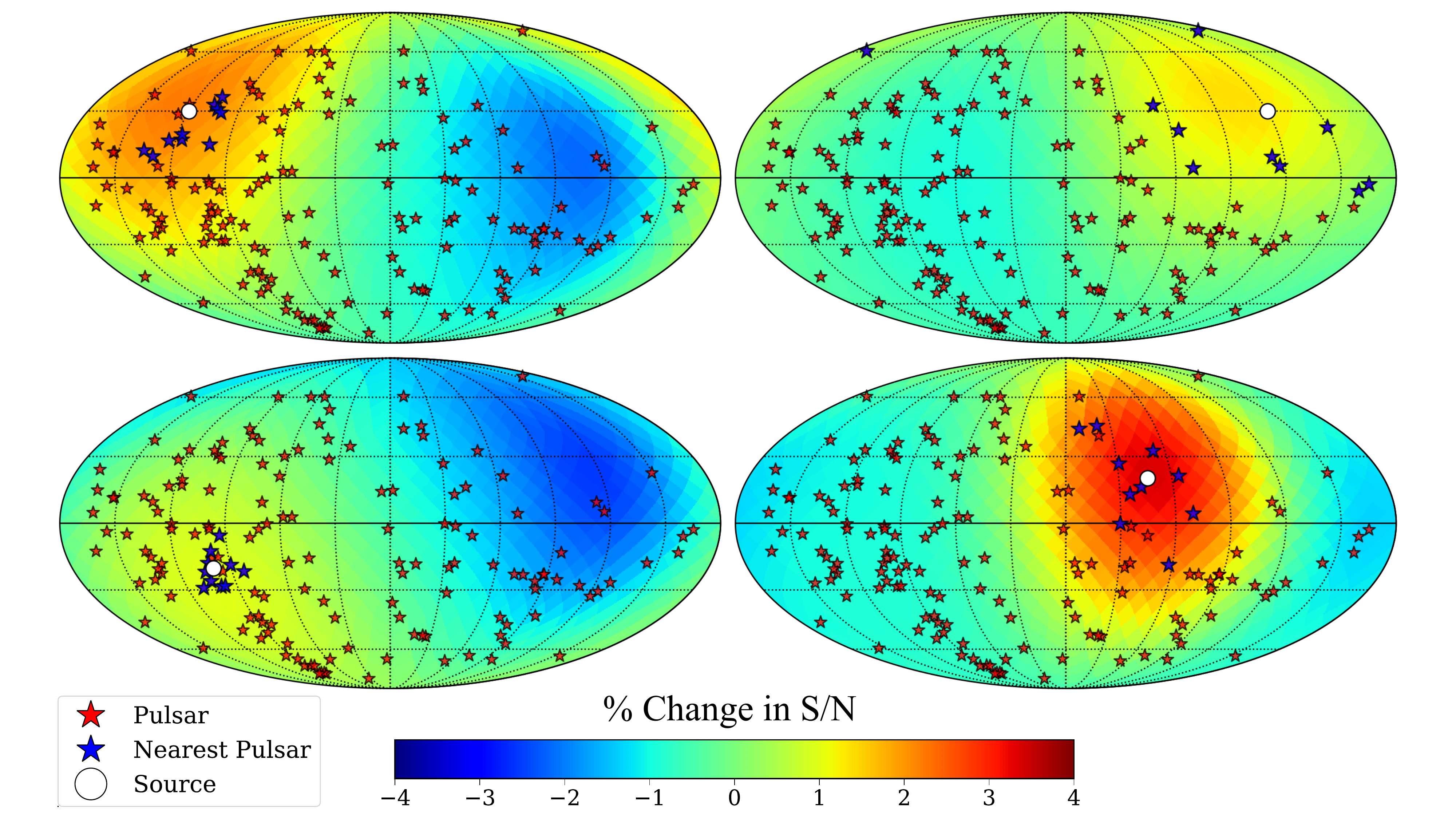}
    \captionsetup{width=.98\linewidth}
    \caption{Sky-targeted campaign. In this alternate campaign, we quadruple the observing cadence of the nearest $10$ pulsars (marked by a blue star) to each source (marked by a white circle) while halving the cadence of the weakest $60$ pulsars (by GWB S/N contribution) such that telescope time is held fixed. The percent changes in S/N are plotted as sky maps for the source frequency of $8$ nHz. The largest percent increase in S/N is $<4\%$. Note that we increase the cadence of the nearest $10$ pulsars to the source at $20$ years so pulsars added to the array past $20$ years are on standard cadence and remain marked in red. We omit the axis labels for legibility, but the sky orientation remains the same as in Figure~\ref{fig:sky_locations}.}
    \label{fig:sky_target_4panel}
\end{figure*}

A ``directional tuning'' of pulsar timing arrays has been previously explored in several works \cite{christy+2014,liu+2023}. The prevailing idea is that observing pulsars more frequently in the approximate (but not exact) direction of a particular source yields higher single-source S/N in that direction on the sky. Using sensitivity curves, we can rapidly test the directional tunability of a PTA by creating a sensitivity sky map, which encodes directional detector sensitivity in discrete, equal-area pixels across the sky. We can add observations or adjust noise for particular pulsars before recomputing these sky maps and calculating the S/N to hypothetical test sources. We consider $4$ test sources of equal strain but differing sky locations to investigate how tunable our simulated detector is to different regions of the sky. We place the nearest $10$ pulsars (which exist in the detector at year $20$) to source of interest on high ($\times 4$) cadence campaigns for the next $10$ years while we put the weakest $60$ pulsars (in terms of GWB S/N contribution excluding those closest to the source) on half cadence campaigns for the same duration. The percent change in source S/N is laid out in $4$ different panels in Figure \ref{fig:sky_target_4panel} corresponding to hypothetical candidates in $4$ distinct sky regions, corresponding to differing pulsar density and hence sky sensitivity.
We note that these fixed observing time targeted campaigns do not increase our S/N very dramatically in the source direction and decrease it by varying amounts in the other directions, which is in agreement with what was found for the ``Targeted'' search in \cite{liu+2023}. We do note that this directional tuning is most effective in the direction with medium but not lowest pulsar density (bottom right) since fewer observations were being taken in that direction to begin with compared to the higher density pulsar regions (top and bottom left) but the nearest pulsars are closer to the source compared to the lowest pulsar density region (top right).

Since telescope time is fixed in this framework, one can think of this as reallocating observations from across the sky, weighted by pulsar density, to a particular region on the sky. So it makes sense that the lower pulsar density regions are affected more since more observations are actually moving to that sky region. This detector in particular is simulated to be IPTA-like in sky-coverage; regional PTAs typically exhibit less homogenous sky-coverage. Therefore, one could expect more directional tunability in those cases. One caveat of our method, however, is that we cannot precisely say how these targeted, high cadence campaigns impact the parameter estimation of the signal. One recent study, \cite{petrov+2024} (which generally agrees with previous studies \cite{sesana+vecchio2010,goldstein+2019}), concludes that S/N is the dominant factor in source sky localization followed by source proximity to pulsars and source chirp mass. While we observe increased S/N to the source and can thereby surmise better sky localization, we cannot full quantify the impact of these campaigns on parameter estimation; moreover, these effects also depend upon GW frequency and source chirp mass in a way that is not explored in this work, but efforts are already underway to incorporate these effects into \hasasia for future studies.

\subsection{Single-source sensitivity gains of a simultaneous GWB fit}\label{sec:double_fit}
\begin{figure}[h]
    \centering
    \includegraphics[width=0.85\textwidth]{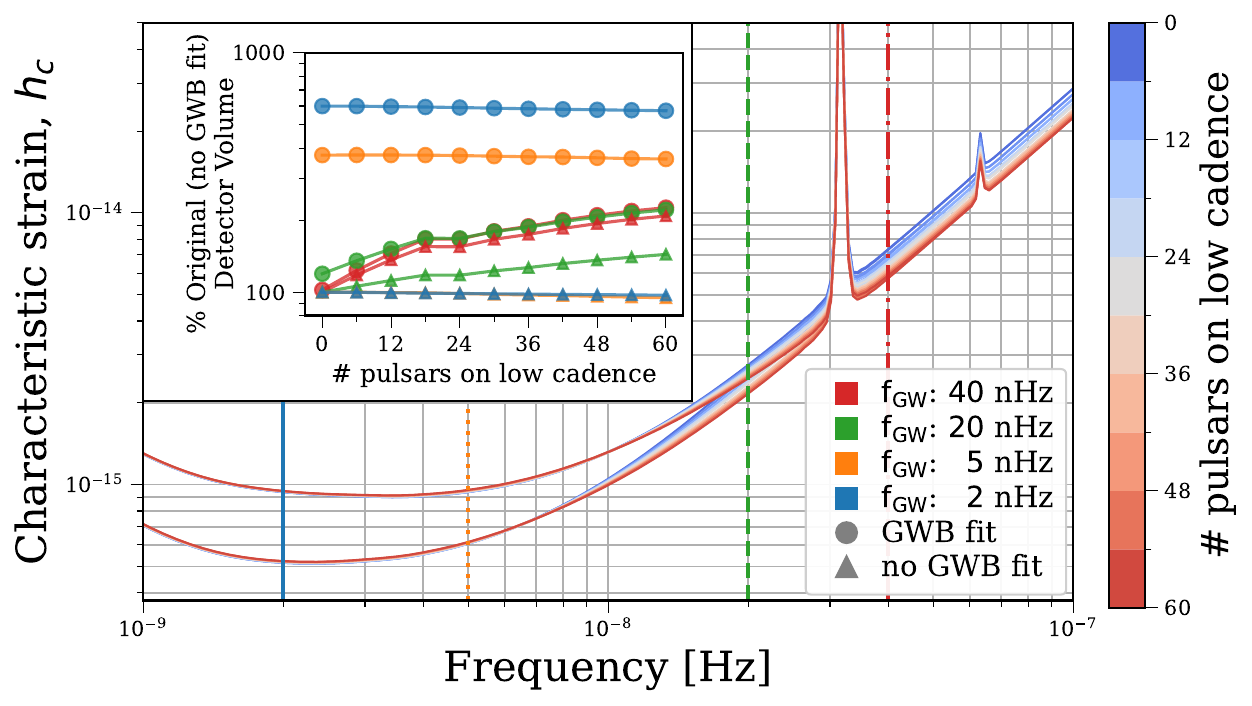}
    \captionsetup{width=0.85\linewidth}
    \caption{The GWB fit. Under the same optimal campaign, we compare single-source sensitivity with a perfect GWB Earth-term fit to no fit at all. The inset plot compares changes in the detector volume with respect to the standard campaign without a GWB Earth-term fit. This is plotted as a function of changes in observing campaign for the cases of with (dots) and without (triangles) a GWB Earth-term fit at the $4$ reference frequencies marked.}
    \label{fig:gwb_subtract_sc}
\end{figure}

The presence of the GWB will necessitate searches for continuous waves (CWs) along with a correlated red noise process in order to model out the power in the Earth-term of the GWB \cite{becsyandcornish2020,ng15singlesource}. Improvements in single-source sensitivity with a simultaneous fit depend upon how well the GWB is fit in the search \cite{ferranti+2024}. We bound the sensitivity gains in this simultaneous fit by considering the limiting cases: a perfect GWB Earth-term simultaneous fit, i.e., only the half of the GWB power in the so-called pulsar term is unmitigated, and no simultaneous fit at all. While our actual sensitivity to single-sources lies somewhere between these two extremes, the potentially drastic difference in sensitivity between these extremes is evident in Figure \ref{fig:gwb_subtract_sc} nonetheless. Our optimal campaign is even more optimal in the case of a simultaneous fit since there is less GWB foreground in the data, which ultimately results in higher sensitivity gains at high frequencies and lower sensitivity losses at low frequencies compared to not performing a simultaneous fit. Notably, the optimal campaign yields sensitivity gains which extend to lower frequencies since the PTA crosses over from being white noise dominated to red noise dominated at lower frequencies with less GWB to contend with. Unsurprisingly, fitting out the power in the GWB Earth-term dramatically improves single-source sensitivity at low frequencies. Recall that the GWB fit is dependent on the number of pulsar pairs and their time-spans, so there is a trade-off between GWB fit and high cadence campaigns in single-source sensitivity.

This hints at the fact that single-source sensitivity as well as the sensitivity gains (and losses) in the optimal campaign and otherwise are dependent on the spectrum of the GWB itself. We explore the relationship between single-source sensitivity and the spectrum of the GWB in the next section. Again, this type of analysis proves trivial in our sensitivity curve framework since all we have to do is inject a differing signal in our noise covariance before recomputing sensitivity.

\subsection{Dependence of Single-Source Sensitivity on GWB Parameters}\label{sec:gwb_dependance}
\begin{figure}[h]
    \centering
    \includegraphics[width=0.85\textwidth]{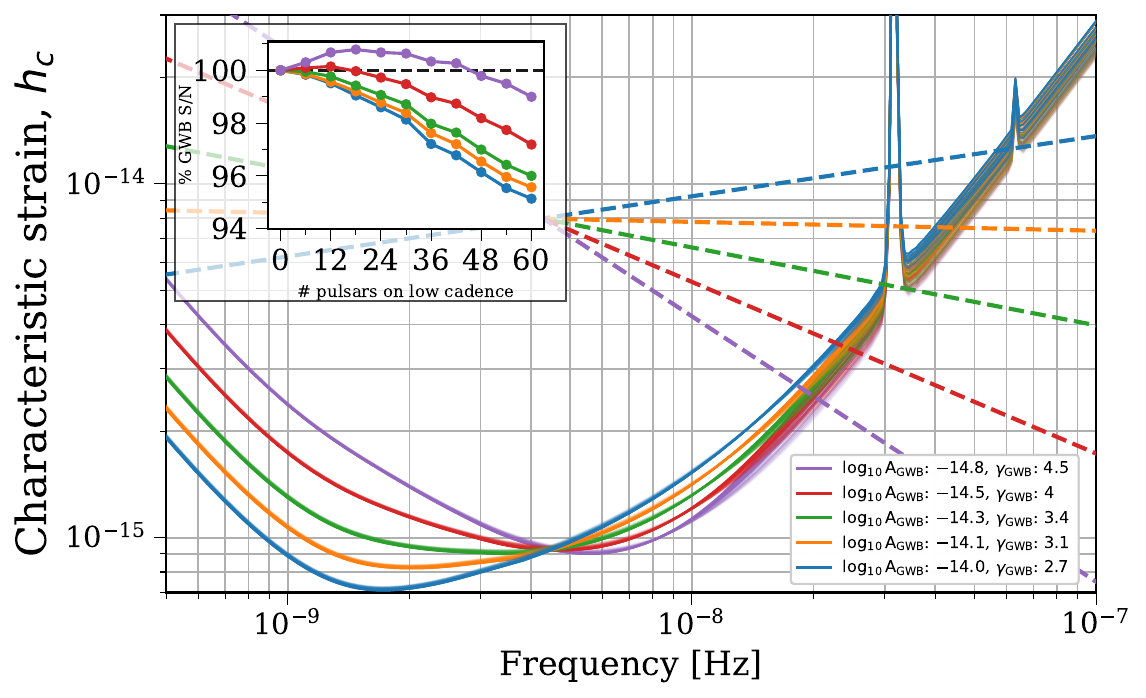}
    \captionsetup{width=0.85\linewidth}
    \caption{Spectral index of the GWB and single-source sensitivity. We simulate single-source sensitivity curves for our detector at $30$ years for various power-law GWB parameter values which are consistent with contemporary PTA observations from \cite{ipta3p+2024}. Single source sensitivity curves  for the optimal campaign (see Section~\ref{sec:opt}) are plotted above and 
    the scaling of each GWB's S/N under the optimal campaign is displayed in the inset plot.}
    \label{fig:varied_gamma_SCs}
\end{figure}

The spectral characterization of the tentative GWB observed in PTAs is currently unconstrained and noise-model dependent \cite{ng15gwb}. Advances in noise modeling techniques as well as more data are expected to improve this in the near future \cite{ng12p5yr_customnoise_inprep,ng15_customnoise_inprep,mpta_dr2,eptadr2_2:noise,inpta_dr1_noise,pptadr3:noise}, but for now, we demonstrate in Figure~\ref{fig:varied_gamma_SCs} how different power-law GWBs affect single-source sensitivity and the efficacy of the optimal observing campaign. We select parameter values of the GWB which are consistent with the joint posterior in \cite{ipta3p+2024} and roughly follow the covariance of the amplitude and spectral index. At high frequencies, the GWB has little impact on single-source sensitivity, and thus the sensitivity curves all align here. Whereas at the lowest frequencies, single-source sensitivity is largely dictated by the power-spectrum of the GWB. We note that the optimal campaign is serendipitously even more optimal for GWBs with steeper spectra than shallower spectra due to the differing frequencies at which the PTA crosses over from being GWB-limited (and red noise-limited) to white noise-limited. In fact, as the inset plot demonstrates for the steepest spectrum ($\gamma_{\rm GWB}=4.5$), the optimal campaign \emph{increases} S/N to the GWB itself at $30$ years for a few pulsars on low cadence. This effect is a result of high cadence campaigns uncovering more GWB from underneath the white noise, thereby boosting the S/N.

\begin{figure}[h]
    \centering
    \includegraphics[width=0.85\textwidth]{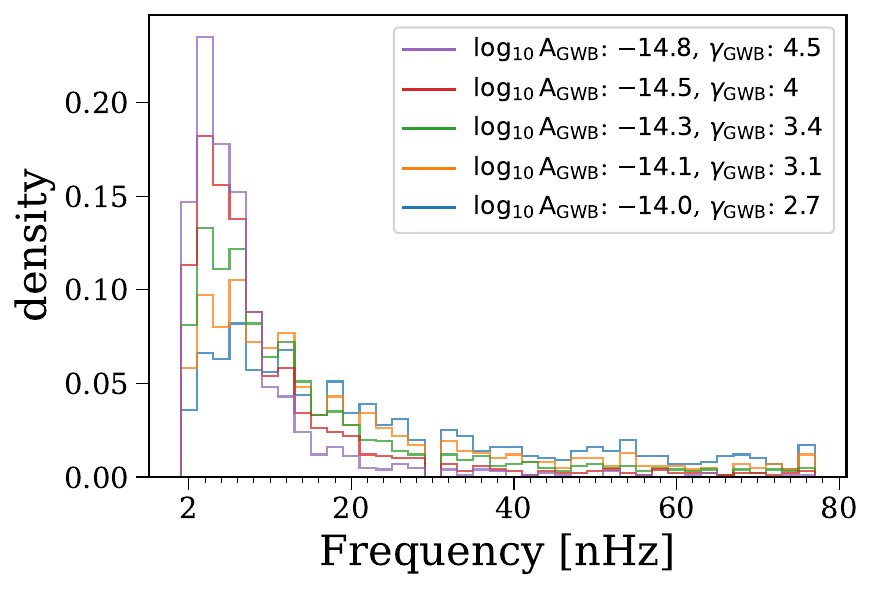}
    \captionsetup{width=0.85\linewidth}
    \caption{Spectral index of the GWB and first single-source detections. A histogram of the frequency bin of the first detected binary across all $1000$ realizations of the SMBBH population. The color encodes injected GWB parameters. Note that the median of first detection shifts to higher frequencies for shallower spectral indices, but the most probable frequency bin remains between the second and fourth frequency bins across different GWBs.}
    \label{fig:varied_gamma_first_detection}
\end{figure}

We can also explore the most probable frequency bin of the first single source detection as a function of the GWB. Using these sensitivity curves and the population simulations (which we adjust for each different GWB), we note the frequency bin of the first binary in each population realization to reach a $3\sigma$ detection. Despite appreciable impacts on single-source sensitivity as seen in Figure~\ref{fig:varied_gamma_SCs}, Figure~\ref{fig:varied_gamma_first_detection} reveals that differing GWBs only slightly impact the most probable frequency bin of the first detection, which is typically in the second or fourth frequency bin ($\sim 4-8$nHz). We note that a shallowed GWB corresponds to a more likely chance to detect the first binary at higher frequencies.

\subsection{Next-Generation Radio Telescope Campaigns}
\begin{figure}[ht]
    \centering
    \includegraphics[width=0.85\textwidth]{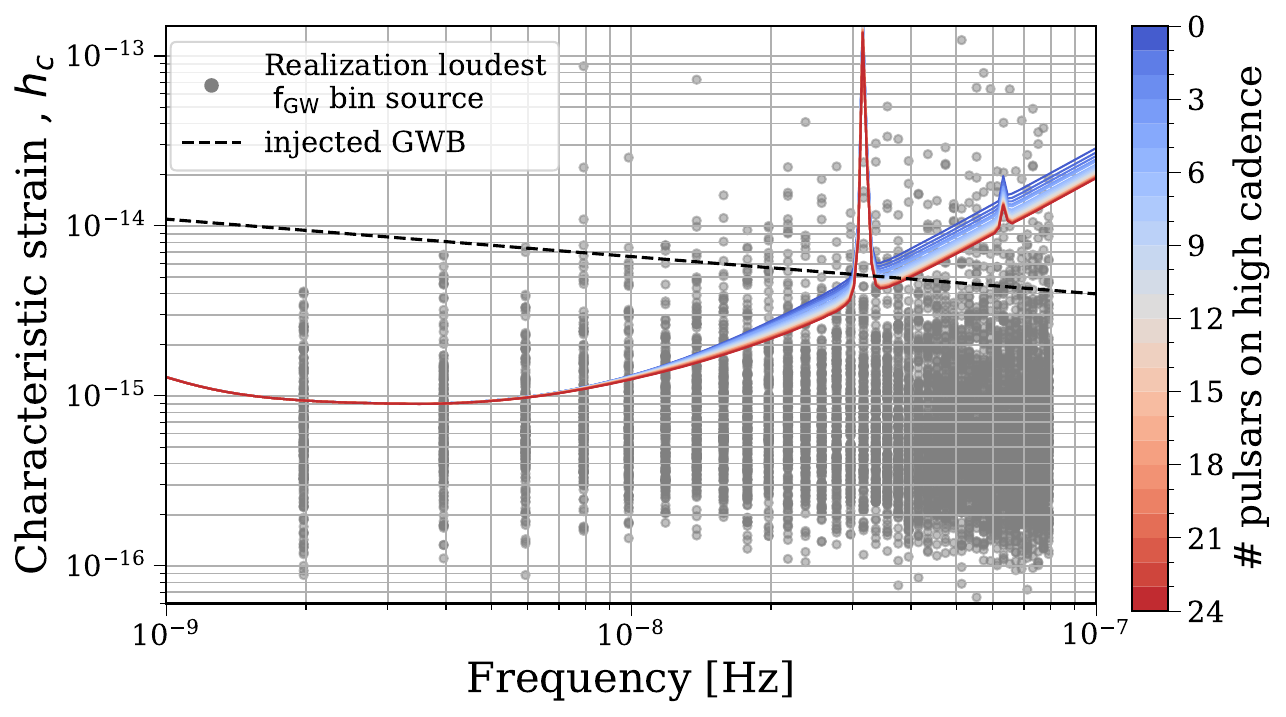}
    \captionsetup{width=0.85\linewidth}
    \caption{Next-generation telescope campaign sensitivity. With the $N$--most sensitive pulsars on a weekly observing cadence with a smaller observational error, we compare single-source strain sensitivity with an optimistic GWB Earth-term fit to the loudest strains in each frequency of $100$ SMBBH population realizations.}
    \label{fig:MTtreatment_sc}
\end{figure}

\begin{figure}[ht]
    \centering
    \includegraphics[width=0.85\textwidth]{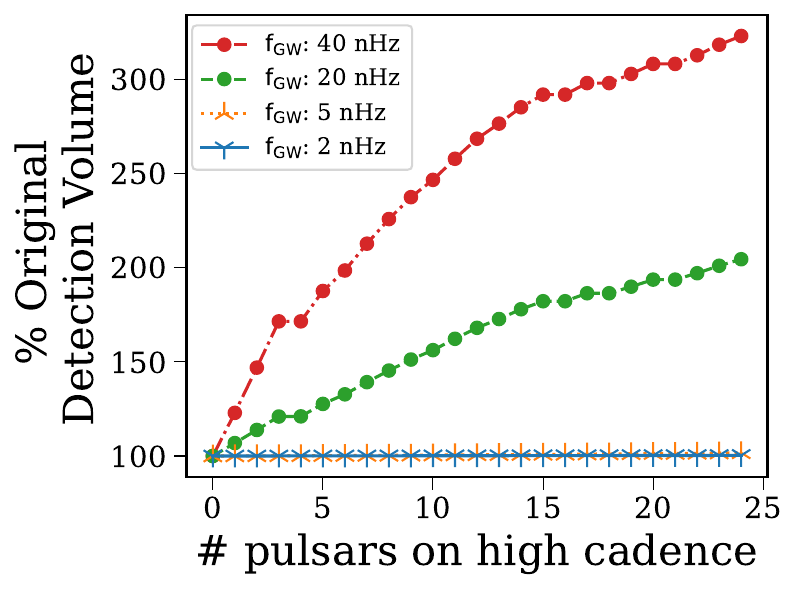}
    \captionsetup{width=0.85\linewidth}
    \caption{ Next-generation telescope campaign detector volume. The change in detector volume is plotted as a function of the number of pulsars on high cadence campaigns with a next-generation telescope.}
    \label{fig:MTtreatment_src_snr}
\end{figure}

\begin{figure}[ht]
    \centering
    \includegraphics[width=0.35\textwidth]{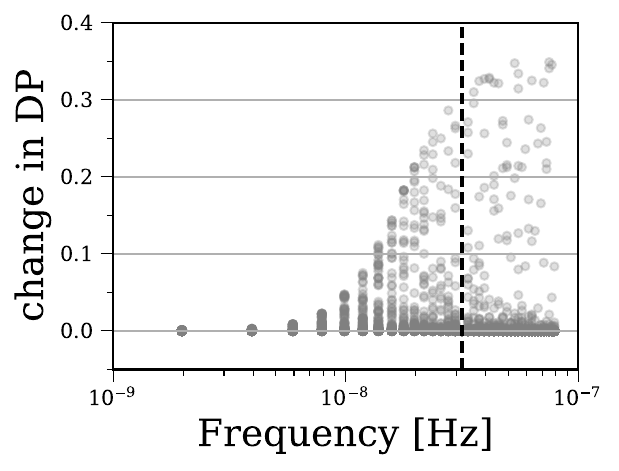}
    \includegraphics[width=0.331\textwidth]{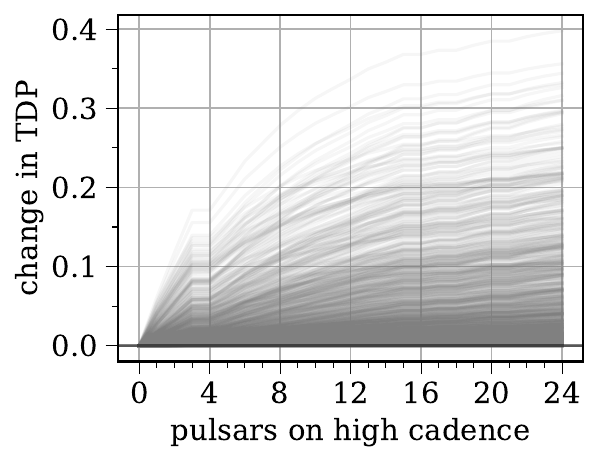} 
    \captionsetup{width=0.85\linewidth}
    \caption{Next-generation telescope campaign detection statistics. (a) The change in DP for each frequency is shown for the case of $24$ pulsars on high cadence campaigns with a next generation telescope. The black line demarcates $1$/yr. (b) We calculate the change in TDP as a function of number of pulsars on a high cadence campaign.}
    \label{fig:MTtreatment_tdp}
\end{figure}

Current and future improvements to radio telescopes hold the promise of significantly reducing white noise in PTA data sets. Here we consider high cadence campaigns conducted by a ``Next-Generation'' radio telescope \emph{without} reducing telescope time. In an IPTA-wide setting, various PTAs have the potential to tune their observations in concert with one another to produce an optimized IPTA data set. For instance, PTAs could use the most sensitive, Next-Generation, radio telescopes such as FAST \cite{FAST2019} or the SKA \cite{ska-proceedings2015} to perform targeted, high cadence campaigns on the most sensitive pulsars. These targeted campaigns are all the more potent as the Next-Generation telescopes achieve observational precision, which is unrivaled by other radio telescopes. To model this scenario, in addition to the standard campaign, we place the most sensitive $N$ pulsars on an observing campaign of $52$ observations per year and also adjust the observational error (RMS) by a factor of $1/10$ to model increased timing precision. Figure~\ref{fig:MTtreatment_sc} shows the sensitivity curves for such a campaign for up to $24$ pulsars on high cadence. The sensitivity gains resemble those of Section~\ref{sec:baseline}, except much more extreme. Figure~\ref{fig:MTtreatment_src_snr} highlights the dramatic increases in detector volume that PTAs could reap from such coordinated campaigns and sensitive telescopes. We note that at high frequencies, the detector volume is more than tripled as compared to the standard campaign. Lastly, we recompute the change in per frequency DP and TPD in Figure~\ref{fig:MTtreatment_tdp} to further demonstrate the benefits of having such an extreme campaign. Since we do not relegate any pulsars to a reduced cadence, we only observe increases in detection probability across all frequencies and this change is even higher than some frequencies as compared with optimal campaign in Section~\ref{sec:opt}. This is a testament to the power of high cadence campaigns and the potential of future radio telescopes to bolster the field of nanohertz GW astronomy.

\subsection{Tuning towards a multi-messenger detection}\label{sec:mma}

Multi-messenger astrophysics (MMA) combines multiple heralds of information to corroborate and enrich the scientific understanding from each source (e.g. \cite{ligo:mma2017}). The Legacy Survey of Space and Time (LSST), which will be carried out by the Vera Rubin Observatory, is expected to give us of order $10^9$ SMBBH candidates from periodic variability in active galactic nuclei (AGN) light curves \cite{lsst2008,witt+2022} due to the unprecedented sensitivity and cadence. However, since the survey will last only for $10$ years, many of these candidates will not extend into the lowest, most sensitive PTA frequency bins. For this reason, it would be beneficial to tune our detector to higher frequencies in the hopes of achieving a multi-messenger detection with the LSST. With the planned time span of $10$ years for the LSST survey, we will not be able to flag any AGN light curves as being potentially periodic below $\sim 10$ nHz, which corresponds to the $5$th frequency bin in our population simulations.

Referring to Figure~\ref{fig:opt_camp_double_fig_rosado} we see that if we focus on frequencies $>10$ nHz the per frequency DP$_i$ all increase for the optimal campaign. This increase in sensitivity at higher frequencies\footnote{Note that anisotropy is predicted to be detected first at higher frequencies due to the larger scatter of binary amplitudes there. \cite{mingarelli+2013_anisotropy, pol+2022}.} would most likely produce a detection of anisotropy sooner, thus revealing the source of the GWB in the nHz band to be SMBBHs rather than cosmological.

\begin{figure}[ht]
    \centering
    \includegraphics[width=0.85\textwidth]{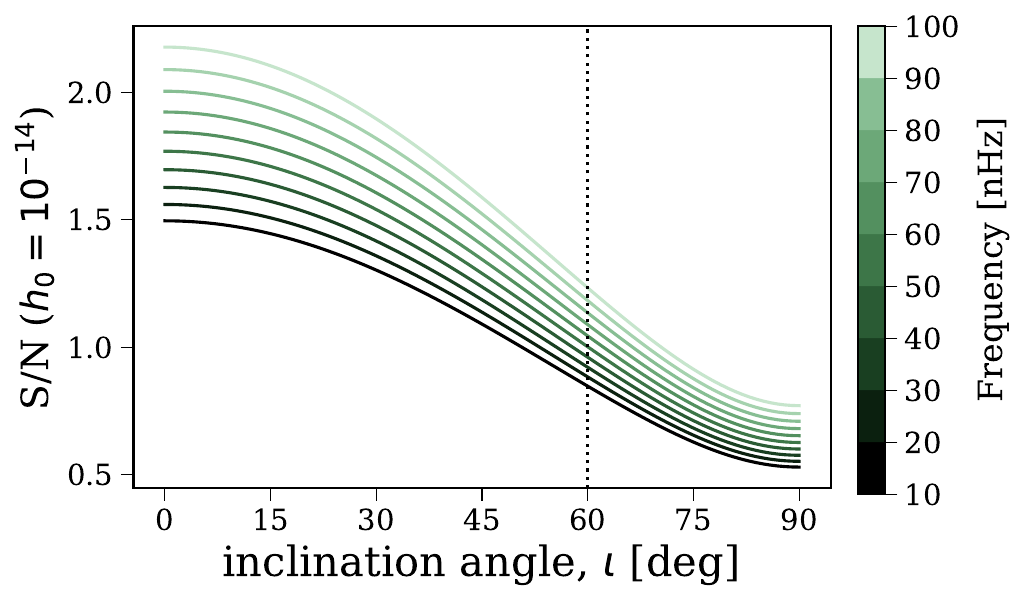}
    \captionsetup{width=0.85\linewidth}
    \caption{Dependence of S/N on inclination angle, $\iota$.}
    \label{fig:snr_vs_inclination}
\end{figure}

Another parameter one might consider with respect to an MMA detection with LSST is the interplay between LSST and PTA sensitivity in terms of binary orientation. For instance, periodicity in AGN is more likely to be discernible from noise for near edge-on ($\iota\sim90^\circ$) binaries due to Doppler boosting and GW self-lensing effects \cite{dorazio+2015,dorazio+2018,kelley+2021lens}, whereas PTAs are more sensitive to face-on binaries ($\iota\sim0$) as can be seen in the sky response functions for GWs (see \ref{sec:appendixA}). Figure~\ref{fig:snr_vs_inclination} compares detection prospects between the optimal PTA binary orientation and edge-on candidate orientations for detection. A ``partial-marginalization'' over only the polarization angle and phase of the S/N is used to do this calculation and is available in \texttt{hasasia}.

\subsection{Sensitivity beyond black hole binaries}
\begin{figure}[ht]
    \centering
    \includegraphics[width=0.85\textwidth]{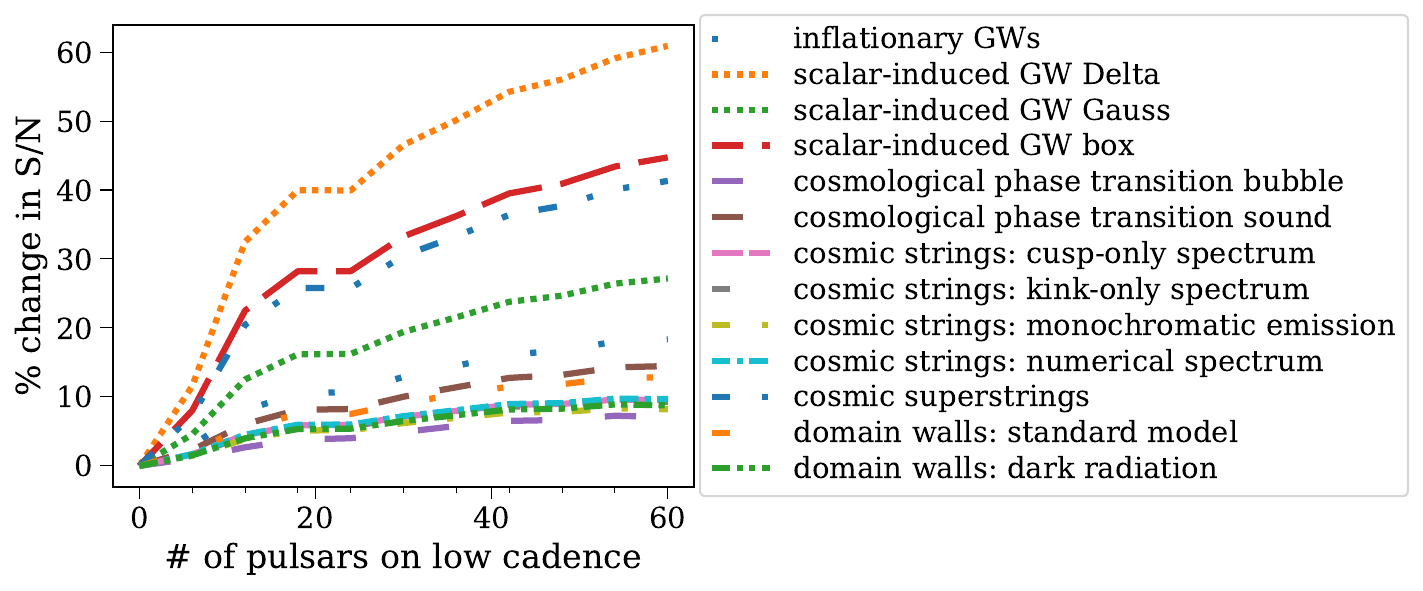}
    \captionsetup{width=0.85\linewidth}
    \caption{New physics sensitivity. The increase in S/N of the optimal campaign considered in Section~\ref{sec:opt} over the standard campaign using the maximum posterior values of several of the new-physics models in \cite{ng15newphysics,ng15newphysicsERRATUM}\label{fig:new_physics_comparison}.}
\end{figure}

Binary black holes are not the only possible source for the emerging stochastic background in PTAs. NANOGrav explored a variety of new-physics models which could produce GWs in the PTA band \cite{ng15newphysics}. Taking the maximum posterior values reported in \cite{ng15newphysics,ng15newphysicsERRATUM}  (see Appendix A, Table $4$), we calculate the change in various S/N values in Figure~\ref{fig:new_physics_comparison} between the standard and optimal campaigns. We find that the optimal observing campaign again yields higher S/N at higher frequencies. In general, the high frequency regime of many of these new-physics signals would more readily allow PTAs to constrain their parameter spaces as they tend to have a shallower effective spectral index than the SMBBH-sourced GWB.

\subsection{Forecasting Parameter Estimation Uncertainties}

Currently the \hasasia framework cannot forecast how the uncertainties of GW parameters will change over time. However, the S/N can be related to the amplitude and spectral index of the GWB \cite{pol+2021astro4cast} and the sky localization of single sources \cite{petrov+2024}.  Pol, et al., 2021\cite{pol+2021astro4cast} found the following empirical scaling relations for the GWB parameter uncertainties as a function of the Hellings-Downs correlated S/N, $\rho_{\rm HD}$,
$\Delta A_{\rm GWB}/ A_{\rm GWB} \approx44\% \times \left (\rho_{\rm HD}/ 3 \right)^{-\frac{17}{25}}$
and
$\Delta \alpha/ \alpha \approx 40\% \times \left (\rho_{\rm HD}/ 3 \right)^{-\frac{17}{20}}$.

In Petrov, et al., 2024 \cite{petrov+2024} they find that the sky localization area in square degrees can be related to the S/N for single sources, $\rho_{\rm ss}$, as $\sim 1/ \rho^2_{\rm ss}$. 

In recent years Fisher matrix analyses have been used to predict the parameter uncertainties for PTAs of the GWB \cite{babak+2024gwb_forecast} and single sources \cite{liu+2023}. These calculations are done in the frequency domain for GWBs \cite{babak+2024gwb_forecast} pointing to the ability to incorporate these types of calculations directly into \hasasia and PTA sensitivity investigations in the future.

\section{Discussion \& Outlook}
\label{sec:discussion}

As pulsar timing arrays enter the detection era, with significance of the GWB anticipated to reach a detection in the coming years, it is natural to focus next on the resolution of anisotropies in the GWB and the detection of individual supermassive binary black holes. This study compares detection statistics for individual sources then demonstrates a powerful method to improve PTA campaigns, using sensitivity curves, which can accurately account for GWB self-noise. Specifically, we find that for our simulated IPTA-like detector,

\begin{enumerate}
    \item Adding high cadence campaigns reduces the power in the white noise, yielding higher sensitivities at $f_{\rm GW}> 10^{-8}$Hz. 
    \item For fixed observing resources, a high cadence (low cadence) campaign on the most (least) sensitive pulsars facilitates high frequency sensitivity gains while yielding minimal sensitivity losses at lower frequencies.
    \item For an IPTA-like detector under fixed observing time constraints, a sky-targeted campaign only minimally boosts sensitivity in that direction. Increasing the sensitivity of the best pulsars, across the sky, gives better gains. 
    \item Simultaneously fitting for a GWB and single-sources provides a large sensitivity boost at low frequencies which depends on the accuracy of the fit.
    \item The parameters of the GWB impacts at what frequencies we expect to see single sources, and impacts where PTAs cross over from being red noise dominated to white noise dominated.
    \item Next generation radio telescopes will drastically reduce the white noise in PTAs thus opening up the possibility for more high frequency candidates.
    \item The sensitivity gains at higher frequencies of the optimal campaign synergizes well with the frequencies possible for an MMA detection with LSST and the expectation that anisotropies will be seen at higher frequencies first.
    \item The optimal observational campaign for PTAs improves sensitivity to various new-physics models at high frequencies, where they could much more easily be disentangled from a SMBBH GWB. 
\end{enumerate}

The use of sensitivity curves skirts the computational cost of simulating many realizations of datasets, and instead allows for an efficient recalculation of the noise in the detector for myriad imaginable scenarios. We demonstrate this method to be versatile in the scenarios it can consider and to be computationally cheap with most simulations being run on a laptop. However, this methodology currently lacks the ability to compute or forecast the parameter estimation of both the GWB and single sources. Future work includes application of this method on a real, reduced PTA dataset with more finely conditioned \texttt{holodeck} SMBBH population simulations as well as an exploration of the interplay between GWB and single-source parameter estimation.


As PTAs approach the detection of gravitational waves from individual supermassive black hole binaries, the successful PTA must account for every iota of sensitivity like a diligent bookkeeper. While additional observational resources are the ideal path to sensitivity gains, realistic constraints leave PTAs with a resource optimization problem. This problem must be addressed sooner rather than later as any changes to a campaign will take time to integrate into the sensitivity.

\section{Acknowledgements}
\label{sec:acknowledgements}

The authors would like to thank Bence B\'ecsy for discussions regarding the various versions of the detection probability. The authors would like to thank Emiko Gardiner and Luke Zoltan Kelley for generously providing the SMBBH population simulations as well as multiple useful conversations regarding them. The authors would like to thank Andrea Mitridate for providing the code used to calculate the new-physics PSDs. Lastly, the authors would like to thank Xavier Siemens for many useful comments throughout the progression of the project. JGB and JSH acknowledge support from NSF CAREER award No.~2339728. JGB is supported in part through NASA and Oregon Space Grant Consortium, cooperative agreement 80NSSC20M0035. JSH acknowledges support from NSF Physic Frontiers Center award No.~2020265 and is supported through an Oregon State University start up fund. JDR also acknowledges support from NSF Physic Frontiers Center award No.~2020265 and start-up funds from the University of Texas Rio Grande Valley.

\textit{Software:} \texttt{hasasia} \cite{hazboun:2019has}, \texttt{holodeck} \cite{holodeck2023}, \texttt{Jupyter} \cite{matplotlib}, \texttt{matplotlib} \cite{matplotlib}, \texttt{numpy} \cite{numpy}, \texttt{scipy} \cite{scipy}

\section{Data and Code Availability}
All the data and code used in this work are available at \url{https://github.com/jeremy-baier/pta_tuning.git}.

\section{References}
\bibliographystyle{iopart-num}
\bibliography{hazgrav}

\appendix

\section{The polarization, inclination and phase averaged $\mathcal{F}_e$ statistic} 
\label{sec:appendixA} 

The $\mathcal{F}$ statistic is a maximum-likelihood detection statistic 
for continuous gravitational-wave searches, originally
developed in the context of ground-based detectors by Jaranowski, Kr\'olak and
Schutz \cite{jaranowski+1998,prix+2009}.
It was later extended to space-based searches by 
\cite{cutler+2005}, and adapted to PTAs as the $\mathcal{F}_e$
statistic\footnote{Note that \hasasia uses the $\mathcal{F}_e$ statistic
instead of the incoherent $\mathcal{F}_p$ statistic~\cite{ellis+2012fp} 
in order to study the sky
position-dependence of single-source sensitivity in the PTA band.} by
\cite{babak+2012}. 

What follows is a summary of the $\mathcal{F}_e$ statistic with enough
information to relate the traditional $\mathcal{F}$ statistic to the statistic
used in \cite{hazboun:2019sc} and originally incorporated into \hasasia, which is averaged over
the gravitational-wave polarization angle $\psi$, inclination angle $\iota$,
and phase $\phi_0$.
Readers interested in more details should refer to the references mentioned above.

As shown in \cite{prix+2009, cutler+2005}, the $\mathcal{F}_e$ statistic
can be written as
\be
2{\cal F}_e
= x_a[{\cal M}^{-1}]^{ab}x_b 
\ee
where
\be
x_a \equiv (x|h_a)\,,
\qquad
{\cal M}_{ab} \equiv (h_a|h_b)\,,
\ee
are defined with respect to the inner product
\be
(x|y)
= 4{\rm Re}\left\{\int_0^\infty{\rm d}f\>
\sum_{I,J}
\tilde x_I(f)
[S^{-1}(f)]_{IJ}\tilde y_J^*(f)\right\}\,.
\ee
Here $[S^{-1}(f)]_{IJ}$ is the inverse matrix
to the matrix of one-sided power spectral densities
$S_{IJ}(f)$ defined via
\be
\langle \tilde n_I(f)\tilde n_J^*(f')\rangle
= \frac{1}{2} S_{IJ}(f)\,\delta(f-f')\,,
\ee
and $S_{I}(f)\equiv S_{II}(f)$.
In addition,
$\tilde x_I(f)$ is the Fourier transform of the time-domain
data $x_I(t)$ for pulsar $I$:
\be
x_I(t) = n_I(t) + \mathcal{A}^ah_{aI}(t)\,,
\ee
where $\mathcal{A}^a$ are amplitude parameters which depend on
$\{h_0,\cos\iota,\psi,\phi_0\}$ via
\be
\begin{aligned}
&{\cal A}^1 = +A_+ \cos 2\psi \cos\phi_0 - A_\times\sin 2\psi\sin\phi_0\,,
\\
&{\cal A}^2 = +A_+ \sin 2\psi \cos\phi_0 + A_\times\cos 2\psi\sin\phi_0\,,
\\
&{\cal A}^3 = -A_+ \cos 2\psi \sin\phi_0 - A_\times\sin 2\psi\cos\phi_0\,,
\\
&{\cal A}^4 = -A_+ \sin 2\psi \sin\phi_0 + A_\times\cos 2\psi\cos\phi_0\,,
\end{aligned}
\ee
with
\be
A_+ \equiv h_0\left(1+\cos^2\iota\right)\,,
\qquad
A_\times \equiv 2h_0 \cos\iota\,,
\label{e:A+Ax}
\ee
and $h_{aI}(t)$ are basis functions defined by
\be
\begin{aligned}
&h_{1I}(t) 
= +\frac{1}{2\pi f_0}\,\sin(2\pi f_0 t)\, F^+_I(\hat k)\,,
\\
&h_{2I}(t) 
= +\frac{1}{2\pi f_0}\,\sin(2\pi f_0 t)\, F^\times_I(\hat k)\,,
\\
&h_{3I}(t) 
= -\frac{1}{2\pi f_0}\,\cos(2\pi f_0 t)\, F^+_I(\hat k)\,,
\\
&h_{4I}(t) 
= -\frac{1}{2\pi f_0}\,\cos(2\pi f_0 t)\, F^\times_I(\hat k)\,.
\end{aligned}
\ee
Here
\be
F^{+}_I(\hat k)\equiv
\frac{1}{2} \frac{\hat p^i_I \hat p^j_I}{1+\hat
p_I\cdot\hat k} e_{ij}^{+}(\hat k)\,,
\\
\qquad
F^{\times}_I(\hat k)\equiv
\frac{1}{2} \frac{\hat p^i_I \hat p^j_I}{1+\hat
p_I\cdot\hat k} e_{ij}^{\times}(\hat k)\,,
\ee
are the $+,\times$ antenna pattern functions for pulsar $I$, 
$\hat p_I^i$ is the unit vector from the Earth to pulsar $I$,
and $e_{ij}^{+,\times}(\hat k)$ are the $+$ and $\times$ polarization
tensors.

It follows that the expectation value of the $\mathcal{F}_e$
statistic with respect to different noise realizations is
\be
\langle 2{\cal F}_e\rangle_{\rm noise} 
=  4+ \rho^2_{\rm MF}({\cal A})\,,
\ee
where
\be
\rho^2_{\rm MF}({\cal A}) \equiv (s|s)
\ee
is the (expected) squared matched-filter signal-to-noise ratio, 
evaluated for the true value of the signal parameters.
Explicitly,
\be
\begin{aligned}
\rho^2_{\rm MF}({\cal A})=\frac{T}{(2\pi f_0)^2}
\bigg[
&\sum_I \frac{\left(F_I^+(\hat k)\right)^2}{S_I(f)}
\left(A_+^2\cos^2 2\psi + A_\times^2\sin^2 2\psi\right)
\\ 
+&\sum_I \frac{\left(F_I^\times(\hat k)\right)^2}{S_I(f)}
\left(A_+^2\sin^2 2\psi + A_\times^2\cos^2 2\psi\right)
\\
+&2\sum_I
\frac{F_I^+(\hat k)F_I^\times(\hat k)}{S_I(f)}
\left(A_+^2- A_\times^2\right)\cos 2\psi \sin 2\psi
\bigg]\; . 
\label{eq:fullsnr}
\end{aligned}
\ee
The above result for $\langle 2{\cal F}_e\rangle_{\rm noise}$
allows us to conclude that $2\mathcal{F}_e$ is
described by a non-central $\chi^2$ distribution with 4 DOFs and
non-centrality parameter $\rho_{\rm MF}^2(\mathcal{A})$ \cite{ellis+2012fp}
\be \label{eq:pdf_fe}
\mathcal{P}_{2\mathcal{F}_e}=\frac{(2\mathcal{F}_e)^{1/2}}{\rho}I_1\left(\rho\sqrt{2\mathcal{F}_e}\right)e^{-\left(\mathcal{F}_e+\frac{1}{2}\rho^2\right)}\;,
\ee
where $I_1(x)$ is the modified Bessel function of the first kind of order $1$.

If instead we average $2\mathcal{F}_e$ over the GW polarization
angle $\psi$, inclination angle $\iota$, and phase $\phi_0$, we
obtain
\be 
\label{eq:Fhas_fap}
\langle 2{\cal F}_e\rangle_{\Omega} 
=  2{\cal F}_0 + \langle \rho^2(h_0)\rangle_{\Omega}\,,
\ee 
where $\mathcal{F}_0$ is the equivalent of $\mathcal{F}_e$ 
but for the null hypothesis (i.e., the absence of a GW signal), and 
$\langle \rho^2(h_0)\rangle_{\Omega}$ is the  
squared signal-to-noise ratio defined in \cite{hazboun:2019sc} and 
used in \hasasia, obtained by averaging over angles and phase,
$\langle\ \rangle_\Omega\equiv \langle \ \rangle_{\cos \iota, \psi, \phi_0}$.
Explicitly,
\be
\begin{aligned}
\langle \rho^2(h_0)\rangle_{\Omega}
&\equiv \langle \rho^2_{\rm MF}({\cal A})\rangle_{\cos \iota, \psi, \phi_0}
\\
&=
\frac{1}{2}\int_{-1}^{1} {\rm d}(\cos\iota)\>
\frac{2}{\pi}\int_{-\pi/4}^{\pi/4} {\rm d}\psi\>
\frac{1}{2\pi}\int_{0}^{2\pi} {\rm d}\phi_0{}\>
\rho^2_{\rm MF}({\cal A})
\\
&= 
\left(\frac{1}{2}\right)\left(\frac{4}{5}\right)
T\frac{h_0^2}{(2\pi f_0)^2}
\sum_I\frac{(F^+_I(\hat k))^2 + (F^\times_I(\hat k))^2}
{S_I(f)}\, ,
\label{e:avg_rho2}
\end{aligned}
\ee
where the factor of $1/2$ in the last line comes from averaging over the
polarization angle $\psi$, and the factor of $4/5$ comes from averaging
over $\cos\iota$.

If we further average (\ref{eq:Fhas_fap})
over noise realizations, we obtain
\be
\langle\langle 2\mathcal{F}_e\rangle_{\Omega} \rangle_{\rm noise} 
=  4 + \langle \rho^2(h_0)\rangle_{\Omega}
= \langle\langle 2\mathcal{F}_e\rangle_{\rm noise}\rangle_{\Omega}\,,
\ee
which shows that noise averaging and angle averaging commute.
With angle averaging the distribution of the 
$\langle{2\mathcal{F}_e}\rangle_{\Omega}$ 
statistic is an \emph{ordinary} $\chi^2$ distribution with
4 DOFs, but shifted to the right by $\langle \rho^2(h_0)\rangle_{\Omega}$
\be \label{eq:pdf_fe_ang_ave}
\mathcal{P}_{\langle 2\mathcal{F}_e\rangle_{\Omega}}=\left(\mathcal{F}_e-\frac{1}{2}\rho^2\right)\exp^{-(\mathcal{F}_e-\frac{1}{2}\rho^2)}\;.
\ee
In the absence of a signal, both of these statistics are 
identical and are described by an ordinary $\chi^2$ distribution
with 4 DOFs. This is the distribution used for the false alarm probability
versus threshold relation.

\begin{figure}[ht]
    \centering
    \includegraphics[width=0.9\textwidth]{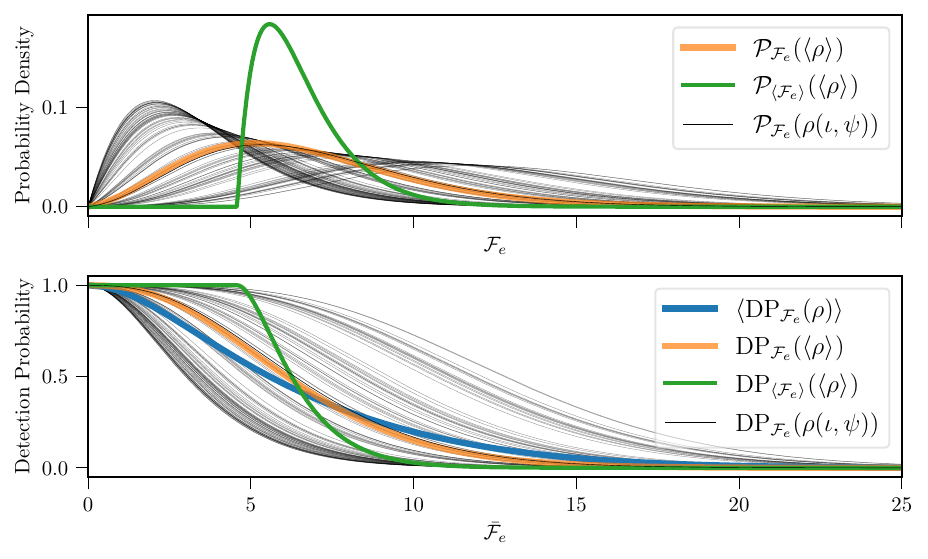}
    \captionsetup{width=0.9\linewidth}
    \caption{Comparison of the probability density functions and detection probabilities for various versions of the \fstat. In the legends above all $\langle\cdot\rangle$ refer to averaging over $\cos \iota$, $\psi$ and $\phi$. These curves are constructed for a signal that has an angle-averaged  $\langle \rho\rangle_{\Omega}\approx 3$. Note particularly in the lower figure that $\langle {\rm DP}_{\mathcal{F}_e}(\rho) \rangle \neq  {\rm DP}_{\mathcal{F}_e}(\langle\rho\rangle) $ }
    \label{fig:DP_compare}
\end{figure}

With these two probability distributions in hand we next demonstrate the differences in the 
detection probabilities (DPs), which depend on \emph{what stage} 
one averages over $\cos \iota$, $\psi$ and $\phi_0$. 
Following \cite{rosado+2015}, the DP gives the probability that a given source will be detected given a threshold for the detection statistic, $\bar{\mathcal{F}_e}$, and is calculated using the probability density function of the \fstat, 
\be\label{eq:dp_P}
\text{DP} = \int_{\bar{\mathcal{F}}_e}^{\infty} \mathcal{P}(\rho, \mathcal{F}_e) d\mathcal{F}_e\; .
\ee
Either Equation~\ref{eq:pdf_fe} or \ref{eq:pdf_fe_ang_ave} can be used for $\mathcal{P}$ in Equation~\ref{eq:dp_P}. The probability density functions and DP functions are shown Figure~\ref{fig:DP_compare}.
The angle-averaged S/N , $\langle \rho\rangle_{\Omega}$, was derived in \cite{hazboun:2019sc} and was the first single source detection statistic incorporated into \hasasia and used to compare the single-source strain threshold to the Bayesian strain upper limits from \cite{agg+19}. For the purposes of forecasting, formalisms that predict the detection probability of single sources,  like the one in Rosado et al.~2015 \cite{rosado+2015}, are preferred because they can be easily related to a false alarm probability and a total detection probability.

The angle-averaged signal to noise, $\langle \rho^2(h_0)\rangle_{\Omega}$, is very closely
related to the $\cal B$-statistic, first developed for ground-based detectors
by Prix and Krishnan \cite{prix+2009} and derived for PTAs by Taylor, Ellis and
Gair \cite{taylor+2014Bstat}. However, in those works all four of the $\cal A$
parameters are marginalized over to form a Bayes factor. As noted in those
works, the $\cal A$ priors are unphysical and lead to slight errors in recovery
and injection studies. Here we instead marginalize over the physical priors of
only the angle parameters in $\bar{\cal A} \equiv \{h_0, \cos\iota, \psi,
\phi_0\}$. This suits the objective of our work here to build sensitivity
curves for single-sources in the PTA band where we would like the freedom to
change $h_0$. While the angle-marginalized signal-to-ratio was used
in previous studies, \hasasia now has the ability to look at sensitivity curves
that depend on $\cos\iota$, $\psi$ and $\phi_0$. In order to more closely mimic a set of simulations, where sources are chosen to have specific values of $\iota$ and $\psi$ and detection probabilities are calculated and, only then, averaged, we use the angle-averaged detection probability, $\left\langle {\rm DP}_{\mathcal{F}_e}(\rho(\iota,\psi)) \right\rangle_{\Omega}$. We are able to calculate the integral over these angles using Python-based numerical integration functionality, retaining the calculational speed necessary for these campaign-tuning investigations.

\end{document}